\documentclass[11pt,letterpaper]{article}

\usepackage{jheppub}
\usepackage{xcolor}
\usepackage{graphicx}
\usepackage{xspace}
\usepackage{wrapfig,enumerate,slashed}
\usepackage{subfigure}
\usepackage{footmisc}
\usepackage{amsmath}
\usepackage{wasysym} 
\usepackage{graphicx}
\usepackage{color}
\usepackage{orcidlink}
\usepackage{comment}
\usepackage{hyperref}
\usepackage{booktabs}
\usepackage{multirow}
\usepackage{cleveref}
\usepackage{physics}
\usepackage{siunitx}
\usepackage{enumitem}
\usepackage[table,dvipsnames]{xcolor}
\usepackage[compat=1.1.0]{tikz-feynman}
\usepackage[normalem]{ulem}
\usepackage{cancel}
\newcommand{\nn}{\nonumber}

\renewcommand{\d}{{\mathrm{d}}}

\newcommand{\Sc}{\ensuremath{S}}
\newcommand{\Fm}{\ensuremath{\Psi}}
\newcommand{\Vc}{\ensuremath{X}}
\newcommand{\dkl}{\ensuremath{\Delta\kappaHHH}}
\newcommand{\mjj}{\ensuremath{m_{jj}}}
\newcommand{\detajj}{\ensuremath{\Delta\eta_{jj}}}
\newcommand{\met}{\ensuremath{E_T^{\rm miss}}}

\newcommand{\MSbar}{\ensuremath{\overline{\scriptstyle \mathrm{MS}}}\xspace}
\newcommand{\fin}{\mathrm{fin}}
\newcommand{\PVAz}{\ensuremath{\boldsymbol{\mathrm{A}}_0}}
\newcommand{\PVBz}{\ensuremath{\boldsymbol{\mathrm{B}}_0}}
\newcommand{\PVDBz}{\ensuremath{\boldsymbol{\mathrm{DB}}_0}}
\newcommand{\PVCz}{\ensuremath{\boldsymbol{\mathrm{C}}_0}}

\newcommand{\cpsi}{\ensuremath{{C_\psi}}}
\newcommand{\cff}{\ensuremath{{C_{FF}}}}
\newcommand{\brinv}{\ensuremath{\mathcal{B}_{\text{inv}}}}
\newcommand{\kappaHHH}{\ensuremath{\kappa_\lambda}\xspace}
\newcommand{\kappaZZH}{\ensuremath{\kappa_{h}}\xspace}

\newcommand{\feynrules}{\textsc{FeynRules}\xspace}
\newcommand{\madgraph}{\textsc{MadGraph5}\xspace}
\newcommand{\pythia}{\textsc{Pythia8}\xspace}
\newcommand{\delphes}{\textsc{Delphes}\xspace}
\newcommand{\matchete}{\textsc{Matchete}\xspace}
\newcommand{\feynarts}{\textsc{FeynArts}\xspace}
\newcommand{\formcalc}{\textsc{FormCalc}\xspace}
\newcommand{\looptools}{\textsc{LoopTools}\xspace}

\definecolor{distinguishable}{HTML}{7A9EC9} % white
\definecolor{confVector}{HTML}{F0C674}      % gold   (Plot9 vector colour)
\definecolor{confFermion}{HTML}{EF9A9A}     % salmon (Plot9 fermion colour)

\crefname{equation}{Eq.}{Eqs.}
\Crefname{equation}{Equation}{Equations}
\crefname{figure}{Fig.}{Figs.}
\Crefname{figure}{Figure}{Figures}
\crefname{table}{Tab.}{Tabs.}
\Crefname{table}{Table}{Tables}

\DeclareSIUnit\barn{b}

\begin{document}
\allowdisplaybreaks
\flushbottom
%%%%%%%%%%%%%%%%%%%%%%%%%%%%%%%%%%%
%\title{Disentangling the Higgs Portal}
\title{A phenomenological profile of Higgs portals at future colliders}
%%%%%%%%%%%%%%%%%%%%%%%%%%%%%%%%%%%
\author[a]{Maximilian Detering\orcidlink{0009-0001-1408-8192},}
\author[b]{Christoph Englert\orcidlink{0000-0003-2201-0667},}
\author[a]{Victor Maura\orcidlink{0000-0003-2930-6356},}
\author[a]{Tevong You\orcidlink{0000-0003-2391-7463}}
%%%%%%%%%%%%%%%%%%%%%%%%%%%%%%%%%%%
\affiliation[a]{Physics Department, King’s College London, Strand, London, WC2R 2LS, United Kingdom}
\affiliation[b]{Department of Physics \& Astronomy, University of Manchester, Oxford Road,\\Manchester M13 9PL, United Kingdom}
%%%%%%%%%%%%%%%%%%%%%%%%%%%%%%%%%%%

\emailAdd{maximilian.detering@kcl.ac.uk}
\emailAdd{christoph.englert@manchester.ac.uk}
\emailAdd{victor.maura\_breick@kcl.ac.uk}
\emailAdd{tevong.you@kcl.ac.uk}
%%%%%%%%%%%%%%%%%%%%%%%%%%%%%%%%%%%

\abstract{Should the Higgs boson be found to couple to a hidden dark sector, it will become necessary to determine the nature of the Higgs portal responsible. Could the minimal, marginal Higgs portal operator be distinguished from a Higgs portal interaction involving a higher-dimensional operator? This is unlikely for Higgs portals with on-shell invisible decays to light dark sectors, but for an off-shell Higgs portal above threshold, where the dark sector is too heavy to contribute to the invisible Higgs decay width, we find that the higher-dimensional Higgs portal operators could in principle be differentiated by combining measurements across the integrated programme of HL-LHC, FCC-ee, and FCC-hh. This illustrates the complementarity of precision and high-energy exploration to phenomenologically profile Higgs portal interactions and probe possible connections between the electroweak scale and a hidden universe.}

%%%%%%%%%%%%%%%%%%%%%%%%%%%%%%%%%%%
\preprint{KCL-PH-TH/2026-23}
\maketitle
%%%%%%%%%%%%%%%%%%%%%%%%%%%%%%%%%%%

%%%%%%%%%%%%%%%%%%%%%%%%%%%%%%%%%%%%%%%%%%%%%%%%%%%%%
\section{Introduction}
\label{sec:intro}
%%%%%%%%%%%%%%%%%%%%%%%%%%%%%%%%%%%%%%%%%%%%%%%%%%%%%

Higgs physics is entering a precision era of its own, with plans for future Higgs factories over the coming decades set to bring the Higgs sector into sharp focus~\cite{EuropeanStrategyGroup:2020pow, P5:2023wyd, CERN-ESU-2025-002}, echoing the electroweak precision programme of a generation ago that transformed our understanding of the gauge sector. The high-luminosity Large Hadron Collider (HL-LHC), with its upgrade currently on the way, together with FCC-ee~\cite{Benedikt:2651299} and FCC-hh~\cite{FCC:2018vvp}, will potentially measure the Higgs couplings, self-coupling, mass and decay widths down to the percent-level precision and better. The continued exploration of the Higgs sector is motivated by the Higgs boson being central to many shortcomings of the Standard Model (SM) and responsible for much of the remaining arbitrariness in its structure. The necessity for this broad programme of fundamental physics includes not only testing for deviations in the Higgs couplings to SM particles but also probing entirely new interactions directly.

Chief among such new interactions is a possible coupling to a hidden sector uncharged under the SM. Hidden sectors of varying complexity could be ubiquitous: almost any SM extension with a SM gauge singlet introduces some portal interaction, if only generated radiatively. The possible portals without a large suppression from an effective field theory (EFT) perspective are, however, limited. The Higgs portal, coupling through the gauge singlet operator $H^\dagger H$, is uniquely positioned as a lowest-dimensional gauge singlet operator\footnote{The other operators of sufficiently low dimension are the hypercharge field strength $B_{\mu\nu}$ via kinetic mixing and the lepton-Higgs combination $LH$ coupling to another fermion, giving rise to the dark photon portal and the neutrino portal, respectively.}; it is the only one that couples to a bilinear of hidden-sector fields rather than mixing with a single one, so it is able to form the most potential interactions with fields of a hidden sector. Investigations of the phenomenological Higgs profile (first pioneered in~\cite{Ellis:1975ap}), therefore, become an indirect phenomenological profile of Higgs portal interactions in the precision Higgs era. The marginal portal requires a hidden bilinear of mass dimension two, and so is available only if the hidden sector contains a light scalar.\footnote{The Proca bilinear $V_\mu V^\mu$ also has mass dimension two, but a massive vector is not renormalisable on its own and requires a UV completion; if the mass arises from a dark Higgs mechanism, the portal is again mediated by the dark Higgs, and in the Stückelberg case by the associated scalar.} A fermion bilinear and dark field strength squared, for example, instead give portals suppressed by a heavy scale~$\Lambda$.

Since we have no direct knowledge of the hidden sector's field content, there is no a priori reason to prefer one portal over another. Singlet scalars coupled through the marginal portal are phenomenologically interesting and well-motivated: they can drive a first-order electroweak phase transition, opening the door to electroweak baryogenesis~\cite{Morrissey:2012db,vandeVis:2025efm}, and they arise in scenarios of neutral naturalness, where twin-Higgs type solutions to the hierarchy problem predict a hidden sector coupled to the SM exclusively through the Higgs~\cite{Chacko:2005pe}. Gauge singlet fermions are motivated by neutrino masses and appear in the same model-building efforts that feature singlet scalars~\cite{Barbieri:2005ri,Barger:2007im}, while hidden gauge sectors are generic in hidden valley~\cite{Foot:1991bp,Strassler:2006im} and composite~\cite{Frigerio:2012uc,Carmona:2015haa} constructions. In these cases, the portal is an effective operator, generated by integrating out a heavier mediator.
Observational evidence for dark matter~\cite{Bertone:2004pz,Cirelli:2024ssz} provides further motivation for the existence of neutral states beyond the SM, though we do not require the hidden sector here to account for the observed dark matter relic abundance.

Higgs portals are particularly hard to probe. With SM gauge interactions of the new states being absent, the hidden sector is not abundantly produced at hadron and lepton colliders, and its only imprint on SM processes is through radiative corrections to Higgs correlation functions. In particular, these shift all linear Higgs couplings in common, so that the portal manifests itself as a universal coupling modification of the Higgs couplings to SM matter. Taking the portal to be the leading interaction of the new states with the SM, which is conservative, the effects of a dynamical singlet are correspondingly restricted. The exploration of Higgs portals therefore pushes the frontier of what we can probe in hidden sector extensions of the SM. But with such a restricted phenomenological profile of Higgs portals, it will be important to examine precisely what any effect implies should a signal appear in future measurements. Could we distinguish the different portal interactions phenomenologically?

In principle, the characteristic energy growth associated with a higher-dimensional operator interaction could disentangle an effective portal from the marginal case. In practice, for an \emph{on-shell} Higgs portal where the hidden sector is light enough for the Higgs boson to decay into, the magnitude of the invisible branching ratio would constrain such effects to be too small to be observable at any feasible future collider. For example, we will see that the scale $\Lambda$ for the fermionic Higgs portal $\Bar{\psi}\psi\, H^\dagger H$ is tied to the Higgs invisible branching ratio $\brinv$ as $\Lambda \sim 6.1\text{ TeV}/(\cpsi \sqrt{\brinv})$. A branching ratio at the percent level with an $\mathcal{O}(1)$ Wilson coefficient then places the EFT cut-off scale above $\sim 60$ TeV, far beyond the partonic energies available at any proposed collider. As we show in \cref{sec:discrimination-light}, this conclusion is not specific to the fermionic portal. The prospects are more optimistic, however, for an \emph{off-shell} Higgs portal: if the (${\mathbb{Z}}_2$-symmetric) dark singlet that the Higgs portal couples to is above the threshold of half the Higgs mass, then it avoids entirely the constraints from the invisible Higgs decay width.

Such an off-shell Higgs portal is much more weakly constrained, with many opportunities for potential phenomenological signatures around the corner. At $pp$ colliders, the hidden states can be produced directly together with a Higgs boson, showing up as missing transverse energy (MET) in Higgs production channels, e.g.\ weak boson fusion (WBF)~\cite{Eboli:2000ze}, $ggh + \text{jet}$~\cite{Goodman:2010yf,Bai:2010hh,Djouadi:2012zc}, $tth$~\cite{Kersevan:2002zj}, $Zh$~\cite{Godbole:2003it,Davoudiasl:2004aj}, and $hh$~\cite{Banerjee:2016nzb}. Indirect effects offer a second set of handles. The portal induces universal coupling shifts in the Higgs couplings to SM matter, most precisely probed by the $Zh$ total cross-section at an $e^+ e^-$ collider~\cite{Craig:2013xia}. It also modifies the Higgs trilinear coupling, accessible in di-Higgs production at the HL-LHC and FCC-hh~\cite{He:2016sqr,Voigt:2017vfz} (and indirectly at FCC-ee~\cite{McCullough:2013rea, Maura:2025rcv, terHoeve:2025omu, Allwicher:2025mvd}). Electroweak precision observables, in particular the $W$ mass, provide a third handle. The FCC-hh contributes both higher partonic energies for the direct WBF search and an improvement of roughly a factor of five on the Higgs self-coupling over the HL-LHC~\cite{FCC:2018byv}.

The purpose of this work is to develop a detailed understanding of the possible discriminating power that future Higgs measurements (and potential anomalies) carry for the interpretation of portal-like interactions. We find that an interplay between various measurements at present and future colliders can give valuable information on the nature of an off-shell Higgs portal. The WBF MET channel is typically the most directly sensitive~\cite{Eboli:2000ze,Craig:2014lda,Ruhdorfer:2019utl,Englert:2020gcp} and hence a potential discovery channel at the HL-LHC. However, WBF measurements at the HL-LHC alone will likely not be sufficient to separate the modified kinematics in differential distributions from higher-dimensional operators. The universal coupling shift measurable at a future Higgs factory likewise carries little information on its own, since a single on-shell observable cannot distinguish the UV origins of a common rescaling. We show that discrimination instead comes from correlated patterns in different observables, from precision measurements of Higgs couplings, self-couplings, and the $W$ mass to the relative rates of WBF MET between HL-LHC and FCC-hh. We find that the combination of HL-LHC, FCC-ee and FCC-hh measurements can then distinguish the phenomenology of the different Higgs portals.

This work is organised as follows. In \cref{sec:portals}, we introduce the different Higgs portals that are the subject of this work. In \cref{sec:pheno}, we discuss their phenomenological signatures and the observables that constrain them, together with the sensitivities projected at present and future colliders. In \cref{sec:discrimination}, we then clarify how complementary measurements across different colliders and observables can discern the nature of a Higgs portal if a related deviation in Higgs data is revealed, treating the on-shell and off-shell regimes in turn. We summarise our findings in \cref{sec:conc}.

%%%%%%%%%%%%%%%%%%%%%%%%%%%%%%%%%%%%%%%%%%%%%%%%%%%%%
\section{Beyond the marginal portal}
\label{sec:portals}
%%%%%%%%%%%%%%%%%%%%%%%%%%%%%%%%%%%%%%%%%%%%%%%%%%%%%
%%%%%%%%%%%%%%%%%%%%%%%%%%%%%%%%%%%%%%%%%%%%%%%%%%%%%
\begin{table}[!t]
    \caption{Higgs portals coupling to hidden sector currents via operators of mass dimension 4, 5, and 6 that are considered in this work. $\Lambda$ is the scale of the corresponding operator coefficient. We do not include any additional symmetry factors in the normalisation of the couplings.}
    \label{tab:higgs-portal-scalar-currents}
    \vspace{0.25cm}
    \centering
    \begin{tabular}{ccll}
        \toprule
        Op. Dim. & Coeff. & Operator & Type \\
        \midrule
        4 & $\lambda_{S}$          & $(H^\dagger H) S S$                                   & Scalar          \\
        5 & $\cpsi/\Lambda$  & $(H^\dagger H) \Bar{\psi}\psi$                        & Fermion           \\
        6 & $\cff/\Lambda^2$               & $(H^\dagger H) F_{\mu\nu} F^{\mu\nu}$                 & Vector  \\
        \bottomrule
    \end{tabular}
\end{table}
%%%%%%%%%%%%%%%%%%%%%%%%%%%%%%%%%%%%%%%%%%%%%%%%%%%%%

The simplest, minimal portal is a marginal dimension-4 interaction between the Higgs field $H$ and a dark sector scalar $S$ of the form
\begin{equation}
    \mathcal{L}^{\text{portal}}_{\text{dim-4}} = - \lambda_S |H|^2 S^2 \, . 
    \label{eq:marginalHiggsportal}
\end{equation}
A real singlet scalar also admits a linear coupling to $H^\dagger H$~\cite{Binoth:1996au,Patt:2006fw,Schabinger:2005ei,OConnell:2006rsp,Barger:2007im}, but we consider throughout only the most conservative $\mathbb{Z}_2$-symmetric case, which renders the hidden scalar stable. The resulting phenomenology is particularly limited, as mixing with the Higgs through a vacuum expectation value is absent and decays of the singlet are forbidden, so that any indirect effect on SM processes is restricted to radiative corrections and the singlets can only be produced in pairs. Being renormalisable, this portal does not necessarily require further UV completion.

The marginal Higgs portal is only one possible scenario. The Higgs boson could also interact with dark sectors through higher-dimensional operators if the mediator is decoupled to heavier scales. There are many possibilities for higher-dimensional Higgs portal operators; however, since their phenomenology mainly depends on the dimensionality of the operator, we study here only one example per dimension for operators of mass-dimension 4, 5, and 6. The Higgs portals we consider are collected in \cref{tab:higgs-portal-scalar-currents}. This representative subset is sufficient for our study to illustrate whether higher-dimensional Higgs portal operators could in principle be distinguished from the marginal Higgs portal. We include only operators built from scalar currents since other current-current operators such as $(H^\dagger i \overset{\leftrightarrow}{D}_\mu H) \Bar{\psi}\gamma^\mu\psi$ imply couplings to the electroweak gauge bosons at tree level and are therefore tightly constrained by electroweak precision measurements, so that their primary phenomenology is not that of a Higgs portal. For simplicity, we also restrict ourselves to CP-even operators at each dimension and only consider single-particle extensions of the SM with one dark sector particle at a time. This approach allows us to isolate the phenomenological impact of specific degrees of freedom for a representative subset of Higgs portal operators and provides a clear mapping between UV-complete theories and their leading-order EFT descriptions.

The dimension-five Higgs portal operator that we consider involves coupling to a hidden sector Dirac fermion $\psi$:
\begin{equation}
    \mathcal{L}^{\text{portal}}_{\text{dim-5}} = \frac{\cpsi}{\Lambda} |H|^2\bar{\psi}\psi .
    \label{eq:dim5higgsportal}
\end{equation}
Such an interaction can, for example, be generated by integrating out a heavy real scalar mediator coupling to both sectors, where the scale of the portal operator is approximately set by the mediator mass. 

At dimension six, a hidden gauge field may couple to the Higgs through its field strength $F_{\mu\nu}$ via the effective interaction
\begin{equation}
    \mathcal{L}^{\text{portal}}_{\text{dim-6}} = \frac{\cff}{\Lambda^2}|H|^2 F_{\mu\nu} F^{\mu\nu} \, .
\end{equation}
This operator can, for example, be generated at one-loop by heavy states charged under the hidden gauge group whose masses derive in part from the Higgs, so that $\Lambda$ is set by their mass scale and the Wilson coefficient $\cff$ is loop-suppressed.

These effective interactions extend the EFT framework of the SM to include dark sectors as light degrees of freedom in the EFT. The Wilson coefficients $\cpsi, \cff$ are dimensionless, and the energy scale $\Lambda$ is related to the EFT cut-off. Other higher-dimensional Higgs portal operators could be included, for example, a dimension-6 derivatively-coupled Higgs portal to dark sector scalars, or even a marginal Higgs portal operator to vectors, but are beyond the scope of our study. Such EFT-like interactions between visible and dark sectors through heavy mediators are generically expected, for example, in hidden valley models \cite{Foot:1991bp,Strassler:2006im} or composite scenarios of dark matter~\cite{Frigerio:2012uc,Carmona:2015haa,Song:2023jqm}, and have been studied in the context of four-fermion portal operators in \cite{Darme:2020ral}.\footnote{We do not consider cancellations of the dimension-6 extended marginal portal against its renormalisable version. These have been motivated in~\cite{Frigerio:2012uc,Carmona:2015haa,Song:2023jqm}, and are particularly relevant for the portal's relation with dark matter~\cite{Balkin:2018tma,Bruggisser:2016nzw,Bruggisser:2016ixa}. Additional collider and electroweak phase-transition implications can be found in~\cite{Anisha:2025nhr}.}

We focus on the case where the dominant effect of a hidden-sector theory is through the portal interaction, and truncate the EFT expansion at the lowest-dimensional portal operator to capture the leading effects only. The numerical implementation of these scenarios is performed using a custom-built \feynrules~\cite{Alloul:2013bka} model file that forms the basis of our quantitative study below, employing {\feynarts /\formcalc /\looptools}~\cite{Hahn:1998yk,Hahn:2000jm,Hahn:2000kx} and {\madgraph}~\cite{Alwall:2014hca}.

%%%%%%%%%%%%%%%%%%%%%%%%%%%%%%%%%%%%%%%%%%%%%%%%%%%%%
\section{Phenomenological signatures and observables}
\label{sec:pheno}
%%%%%%%%%%%%%%%%%%%%%%%%%%%%%%%%%%%%%%%%%%%%%%%%%%%%%

%%%%%%%%%%%%%%%%%%%%%%%%%%%%%%%%%%%%%%%%%%%%%%%%%%%%%
\begin{table}[!t]
\centering
\caption{Observables used in this work and the projected sensitivities assumed
throughout. WBF entries are yields, and their sensitivity is set by the expected 
signal and background rather than by a single fractional uncertainty.}
\label{tab:obs-summary}
\vspace{0.25cm}
\begin{tabular}{lllc}
\toprule
Observable & Machine & Projected $1\sigma$ sensitivity & Ref. \\
\midrule
$pp \to jj + \met$ (WBF)        & HL-LHC, \SI{3}{\per\atto\barn}   & --- & --- \\
$pp \to jj + \met$ (WBF)        & FCC-hh, \SI{30}{\per\atto\barn}  & --- & ---\\
$\brinv$& HL-LHC & $4\%$ & \cite{CMS:2018tip} \\
$\brinv$& FCC-ee & $0.05\%$ & \cite{selvaggi_2025_n2emg-43f06} \\
\dkl{} (di-Higgs)               & HL-LHC                           & $0.275$ & \cite{ATLAS:2025eii}\\
\dkl{} (di-Higgs)               & FCC-hh                           & $0.05$ & \cite{taliercio_2025_5mqfv-xnd34}\\
$\Delta\sigma/\sigma\,(e^+e^-\to Zh)$ & FCC-ee, \SI{240}{\GeV}     & $2.1\times10^{-3}$ & \cite{de_blas_future_2019}\\
$\Delta m_W/m_W$                & FCC-ee, $Z$ pole                 & $4\times10^{-6}$ & \cite{FCC:2025lpp,deBlas:2025gyz}\\
\bottomrule
\end{tabular}
\end{table}
%%%%%%%%%%%%%%%%%%%%%%%%%%%%%%%%%%%%%%%%%%%%%%%%%%%%%

%%%%%%%%%%%%%%%%%%%%%%%%%%%%%%%%%%%%%%%%%%%%%%%%%%%%%%%%%%%%%%%%%%%%%%%%%%%%%%%%%

We review here the observable signatures used in the comparative analysis of \cref{sec:discrimination}. \Cref{tab:obs-summary} collects the channels and the projected precision
assumed for each. We begin with invisible decays, which dominate the on-shell regime and constrain the portal couplings directly. For heavier singlets in the off-shell regime, where invisible decays of on-shell Higgs bosons are absent, direct production in weak boson fusion (WBF) $pp\to jj(h^*\to\slashed{E}_T)$ provides the leading sensitivity at hadron colliders~\cite{Craig:2014lda,Ruhdorfer:2019utl}. Indirect effects provide complementary information about a Higgs portal. Universal shifts of the Higgs couplings can be constrained in on-shell Higgs boson measurements~\cite{Craig:2013xia}. In particular, at a precision $e^+e^-$ collider like the FCC-ee, the projected sensitivity of the associated $Zh$ cross section enables a detailed study of Higgs properties~\cite{selvaggi_2025_n2emg-43f06} with great indirect power. Electroweak precision observables provide a further indirect handle, most notably the $W$ boson mass, which receives a contribution from the portal through the same virtual effects~\cite{Englert:2020gcp}. The distinction matters for what follows. The direct rate is essentially a counting experiment, whereas the indirect observables depend on the mass and spin of the hidden state and can take either sign, so they retain information about which portal is responsible. An off-shell Higgs propagation also shapes $gg\to hh$ production in the guise of a modified $hh$ threshold behaviour (typically framed as a measurement of the trilinear Higgs boson coupling)~\cite{Baur:2002rb,Baur:2002qd,Dolan:2012rv}. Di-Higgs production is therefore another relevant process to study portal-type interactions~\cite{He:2016sqr,Voigt:2017vfz,Englert:2019eyl}. 

Furthermore, new physics that couples to the Higgs portal can potentially distort the tails of distributions by perturbatively modifying unitarity-controlled cancellations present in the SM below the new physics scale $\Lambda$. HL-LHC and FCC-hh, due to their extensive coverage in partonic centre-of-mass energy, are particularly well-suited to look for such enhancements. Well-motivated processes here are $gg \to VV$ processes (related to the unitarisation of $t\bar t \to VV$ scattering)~\cite{Kauer:2012hd} and the production of four top quarks (probing unitarisation in massive fermion scattering $t\bar t \to t \bar t$). We have investigated these channels but find them to be uncompetitive with WBF throughout the parameter space considered so we do not discuss them further.

%%%%%%%%%%%%%%%%%%%%%%%%%%%%%%%%%%%%%%%%%%%%%%%%%%%%%
\subsection{Invisible Higgs decays}
\label{sec:decays}
%%%%%%%%%%%%%%%%%%%%%%%%%%%%%%%%%%%%%%%%%%%%%%%%%%%%%
The phenomenology of the Higgs portal interactions most crucially depends on whether the on-shell dynamics of the 125 GeV Higgs boson is modified. The sensitivity obtained from Higgs signal strength measurements sets tight constraints on kinematically unsuppressed decays of the 125 GeV Higgs boson to invisible final states. The most up-to-date measurements of the Higgs total decay width imply $\Gamma_{h} = 3.7^{+1.9}_{-1.4}\,\text{MeV}$~\cite{ParticleDataGroup:2024cfk}, in agreement with the Standard Model prediction of $\SI{4.1}{MeV}$~\cite{LHCHiggsCrossSectionWorkingGroup:2011wcg}. A recent determination including electroweak precision constraints obtains $\Gamma_{h} = 4.36^{+0.63}_{-0.50} \, \text{MeV}$~\cite{Fischer:2026bka}. Meanwhile, an upper bound on the invisible Higgs branching ratio has been placed at $\num{10.7}\%$ at 95\%~C.L.~\cite{ParticleDataGroup:2024cfk,CMS:2022ley,ATLAS:2023dnm,ATLAS:2023tkt}, while the SM prediction for the Higgs invisible branching ratio (neutrino final state) is $<10^{-5}$~\cite{LHCHiggsCrossSectionWorkingGroup:2011wcg}. Any observation above this level is therefore a smoking gun for new physics involving hidden sectors. The HL-LHC is expected to improve the bound to about $4\%$~\cite{CMS:2018tip}, and future colliders like the FCC-ee have a prospective sensitivity at the sub-percent level~\cite{selvaggi_2025_n2emg-43f06}, owing to the clean environment and high statistics of $Zh$ and $WW \rightarrow h$ events.

We compute the Higgs decay width for scalar, fermionic and vector Higgs portals at tree level to relate the potentially observed invisible Higgs decay width to couplings and Wilson coefficients of the above operators (see \cref{sec:decay_widths} for the full expressions). Considering each portal individually, the couplings of each portal operator are then simply related to an observed invisible branching ratio of the Higgs, $\brinv$. The expressions below hold in the limit $m \ll m_h/2$, where the full phase space is open, and the decay width for a given coupling is largest. Non-observation of an invisible branching ratio therefore excludes the largest range of couplings in this limit, and the resulting bounds are conservative for heavier hidden states. The portal couplings are constrained as
\begin{equation}
\begin{split}
    \lambda_S & 
    \sim \sqrt{\frac{8\pi m_h \Gamma_h^S}{v^2}} \approx 0.014 \times \sqrt{\brinv} \, , \\
    \frac{\cpsi}{\Lambda} & \sim \sqrt{\frac{16\pi \Gamma_h^\psi}{m_h v^2}} \approx \frac{1}{\SI{6.1}{TeV}} \times \sqrt{\brinv} \, , \\
    \frac{\cff}{\Lambda^2} & \sim \sqrt{\frac{2 \pi \Gamma_h^X}{v^2 m_h^3}} \approx \frac{1}{(\SI{1.5}{TeV})^2} \times \sqrt{\brinv}\, .
\end{split}
\label{eq:invBR}
\end{equation}
%

%%%%%%%%%%%%%%%%%%%%%%%%%%%%%%%%%%%%%%%%%%%%%%%%%%%%%
\subsection{Direct production via weak boson fusion}
\label{sec:wbf}
%%%%%%%%%%%%%%%%%%%%%%%%%%%%%%%%%%%%%%%%%%%%%%%%%%%%%
The singlets in an off-shell Higgs portal can still be produced directly, through an off-shell Higgs produced in WBF. The signal is $pp\to jj(h^\ast\to \slashed{E}_T)$, where the missing energy is carried by the pair of dark sector particles, and the irreducible background is $pp \to V+jj$ with $V$ decaying to neutrinos or to an unobserved lepton. We follow closely the approach of~\cite{Ruhdorfer:2019utl}, which extends and improves the treatment in~\cite{Craig:2014lda}. Although gluon fusion has a larger cross section, weak boson fusion has a cleaner final state and leads in sensitivity, so we focus exclusively on this channel, noting that further discriminating power may be achievable by including additional channels. The observable we use is the expected event count in a signal region defined by the WBF topology. Specifically, the signal region is characterised by two forward tagging jets separated by a large rapidity gap together with a hard missing energy requirement. Signal and background samples are generated at leading order with \madgraph~\cite{Alwall:2014hca} using our own \feynrules~\cite{Alloul:2013bka} model, showered with \pythia and passed through \delphes. The full event selection and cutflows are given in \cref{sec:wbf-appendix}.

The three portals differ in this production channel through the momentum dependence of the production vertex. The higher-dimensional operators grow with the partonic centre-of-mass energy relative to the marginal portal, associated with harder jets and a harder missing energy spectrum for the same total rate. At HL-LHC energies, this difference is mild, and we find that it does not by itself separate the portals. The effect is far more pronounced at FCC-hh, whose larger partonic energies probe the region where the operator dimension matters: for parameter points that are discoverable at HL-LHC, the ratio of FCC-hh to HL-LHC yields differs markedly between the marginal and higher-dimensional portals. It is this energy leverage, rather than the HL-LHC measurement alone, that makes WBF a useful discriminator in \cref{sec:discrimination}.

We use the inclusive rate in the signal region rather than a shape fit. A shape-based analysis could recover some additional discrimination, but would also require a consistent treatment of systematic uncertainties. For the large backgrounds here, this would in many cases render the measurement systematics limited. We therefore take the one-dimensional inclusive rate assuming statistical uncertainties only instead of using differential distributions, which we leave to a more complete experimental WBF exploration in future FCC-hh studies.  

How the inclusive signal rate scales with the parameters of a given portal is defined by its mass dimension. In the heavy-mass regime, this gives the approximate partonic scaling $\hat\sigma\sim\{\hat s^{-1},\hat s^0,\hat s^{+1}\}$ for the scalar, fermion, and vector portals, respectively, where $\hat s$ is the partonic energy probed in the exotics' production. Combined with the effective WBF luminosity, this translates into the approximate mass scaling: 
\begin{equation}
    S \propto 
    \begin{cases}
        \lambda_S^2M^{-4}, & \text{scalar}, \\
        (\cpsi)^2 M^{-2}, & \text{fermion}, \\
        \cff^2  M^{0},  & \text{vector} ,
    \end{cases}
\end{equation}
which works very well at the FCC-hh and gives a rough estimate at the HL-LHC.  

For orientation, at a hidden sector benchmark mass of \SI{100}{\GeV} and for portal couplings chosen to give a comparable signal, $\lambda_S = 1$, $\cpsi/\Lambda=\SI{5}{\per\TeV}$ and $\cff/\Lambda^2 = \SI{5}{\per\square\TeV}$, the HL-LHC (FCC-hh) signal region yields $S/\sqrt{B}=1.2\,(31)$, $1.9\,(74)$ and $1.3\,(150)$ for the scalar, fermion and vector portals respectively at 3 (30) \si{\per\atto\barn}.

%%%%%%%%%%%%%%%%%%%%%%%%%%%%%%%%%%%%%%%%%%%%%%%%%%%%%
\subsection{\texorpdfstring{Associated $Zh$ production at FCC-ee}{Associated Zh production at FCC-ee}}
\label{sec:eezh}
%%%%%%%%%%%%%%%%%%%%%%%%%%%%%%%%%%%%%%%%%%%%%%%%%%%%%
Any field coupling to the Higgs also modifies the Higgs couplings through radiative corrections. This is because of the contribution to the Higgs wavefunction renormalisation, leading to universal coupling modifications because of the field redefinition $h \to Z_h^{1/2} h = (1 + \frac{1}{2} \delta {Z_h} + \dots)h$ after renormalisation. These universal shifts modify all linear Higgs couplings, such as the Yukawa couplings to the fermions, and can be most precisely probed through the associated production cross section $\sigma(e^+e^- \to Zh)$ at an $e^+e^-$ Higgs factory. Such coupling modifications are commonly described in the $\kappa$-framework, which captures the relative deviation of the measured on-shell couplings from the SM prediction, for instance $\kappa_{ZZh} = {g_{ZZh}}/{g^\text{SM}_{ZZh}}$ for the $ZZh$ vertex, with $\Delta\kappa \equiv \kappa - 1$. The universal Higgs coupling modifications are given in terms of the Higgs wavefunction renormalisation
\begin{equation}
    \kappa_{ZZh} = \kappa_{WWh} = \kappa_{t} = \kappa_{b} = \dots = 1 + \frac{1}{2} \delta {Z_h^\text{finite}},
\end{equation}
where `finite' refers to the UV-finite remainder after renormalisation.\footnote{Other Wilson coefficients, including those generated radiatively or through RGE-mixing, may contribute to a subset of the coupling modifiers and thereby modify this relation. We are therefore implicitly assuming that all other Wilson coefficients vanish at the scale at which the coupling modifiers are defined.} We compute $\delta Z_h$ at one-loop for each portal, with details of the renormalisation procedure given in \cref{sec:renormalisation}). As all coupling modifications are aligned, we denote them collectively as $\kappa_i\equiv\kappaZZH$ in what follows. Since the associated production cross section scales as the square of the $ZZh$ coupling, the mediator's contribution enters the rate as
\begin{equation}
  \frac{\Delta\sigma}{\sigma}(e^+e^- \to Zh) \;\simeq\; 2\,\Delta\kappa_{ZZh}
\end{equation}
to linear order. The most precise determination will come from the Higgs $b\bar{b}$-channel with a relative projected uncertainty of ${\Delta \sigma}/{\sigma} = 0.21\%$ at FCC-ee, extracted from \cite{de_blas_future_2019} and rescaled to match the updated integrated luminosity from \cite{FCC:2025lpp}. 

%%%%%%%%%%%%%%%%%%%%%%%%%%%%%%%%%%%%%%%%%%%%%%%%%%%%%
\subsection{Higgs pair production as a probe of the trilinear Higgs coupling}
\label{sec:kappalambda}
%%%%%%%%%%%%%%%%%%%%%%%%%%%%%%%%%%%%%%%%%%%%%%%%%%%%%
Beyond modifications of the Higgs two-point function, a Higgs portal also modifies the Higgs trilinear coupling, which is accessible, e.g., through di-Higgs production. At one-loop, the Higgs three-point function receives additional one-particle irreducible (1PI) contributions that depend on the mass and spin of the underlying singlet, so that this observable carries information about the underlying new physics that a universal coupling shift alone does not.

The modification of the Higgs trilinear coupling can be captured in the $\kappa$-framework through $\kappaHHH \equiv g_{hhh}/g_{hhh}^\text{SM}$, with the subtlety that, already starting at NLO, $\kappaHHH$ is in general scheme- and gauge-dependent (unless viewed as a Higgs Effective Field Theory (HEFT) parameter \cite{Longhitano:1980iz,Feruglio:1992wf,Buchalla:2017jlu,Brivio:2013pma,Herrero:2021iqt}). Explicitly, the Higgs coupling modification is defined as \cite{Bahl:2023eau,He:2016sqr}
\begin{equation}
    \kappaHHH = 1 + \frac{\Re\hat{\Gamma}_{hhh} - \Re\hat{\Gamma}_{hhh}^\text{SM}}{\Gamma_{hhh}^\text{tree}}\; ,
    \label{eq:kappa-lambda}
\end{equation}
where $\hat{\Gamma}_i$ are the renormalised vertex functions and $\Gamma_{hhh}^\mathrm{tree} = {3m_h^2}/{v}$. We compute $\hat{\Gamma}_{hhh}$ at one-loop order for each portal, with details given in \cref{sec:renormalisation}. Furthermore, note that the 1PI vertex three-point function depends on the external momenta. For definiteness, we choose the kinematical point $p_1^2 = p_2^2 = m_h^2, p_{12}^2 = 4 m_h^2$, corresponding to the threshold for di-Higgs production. The vertex function can be complex through absorptive parts, and we identify $\kappaHHH$ as the real part of the coupling modification only. The virtual corrections will typically induce other operator structures, yet the $hh$ threshold is known to be particularly sensitive to $\kappaHHH$ modifications, and momentum-dependent effects are generally found to be moderate when they arise from higher-order corrections~\cite{Anisha:2024ljc}. We therefore report our results using this parameter as a proxy for the full cross section for comparability (see Refs.~\cite{Baur:2002rb,Baur:2002qd,Dolan:2012rv} for related discussions).

The difference in precision between colliders matters for our analysis. Two portals may give the same signal at HL-LHC while being distinguishable at FCC-hh, so we keep the two machines separate.
The HL-LHC is projected to measure the Higgs coupling deviation with a symmetrised uncertainty of $\sigma(\dkl) = 0.275$ from di-Higgs production at \SI{3}{\per\atto\barn}~\cite{ATLAS:2025eii}, which could be indirectly improved at FCC-ee~\cite{Maura:2025rcv, terHoeve:2025omu, Allwicher:2025mvd}, while FCC-hh is projected to reach an uncertainty of $\sigma(\dkl) = 0.05$ \cite{taliercio_2025_5mqfv-xnd34}. As measurements of the same theoretical quantity, they will be almost perfectly correlated, and the FCC-hh entry strictly dominates.

%%%%%%%%%%%%%%%%%%%%%%%%%%%%%%%%%%%%%%%%%%%%%%%%%%%%%
\subsection{\texorpdfstring{The $W$ boson mass}{The W boson mass}}
\label{sec:mw}
%%%%%%%%%%%%%%%%%%%%%%%%%%%%%%%%%%%%%%%%%%%%%%%%%%%%%
Indirect effects of states coupled to the Higgs also propagate into other electroweak precision observables such as the $W$ mass. The contribution of a Higgs portal is a two-loop effect; a full calculation lies beyond the scope of this work. Instead, we capture the leading contribution to $m_W$ by integrating out the singlet in an effective approach that is approximately valid for heavy masses. The matching contributes to $C_{H\square}$ at one-loop, which in turn runs into $C_{HD}$ (that maps onto the Peskin-Takeuchi $T$ parameter~\cite{peskin_estimation_1992}), corresponding to a shift in $m_W$. The matching is performed using \matchete~\cite{Fuentes-Martin:2022jrf} and we find
\begin{equation}
      C_{H\square}^S = -\frac{1}{16 \pi^2} \frac{\lambda_S^2}{6 m_S^2} \,, \qquad
      C_{H\square}^\psi = \frac{1}{16 \pi^2} \frac{8}{3}\frac{(\cpsi)^2}{\Lambda^2} \, ,
\end{equation}
for the Higgs portal to dark sector scalars and fermions, respectively. Using the anomalous dimension $\gamma(C_{HD},C_{H\square}) = (20 g'^2)/3$ \cite{Alonso:2013hga} and running at leading logarithmic order from the matching scale down to $m_W$, the resulting mass shift is
\begin{equation}
    \frac{\Delta m_W}{m_W} = \SI{-0.12}{\giga\electronvolt^2} \frac{\lambda_S^2}{ m_S^2}\log{\frac{m_S}{m_Z}} \,, \qquad
    \frac{\Delta m_W}{m_W} = \SI{1.96}{\giga\electronvolt^2} \,\frac{\left(\cpsi\right)^2}{\Lambda^2}\log{\frac{m_\psi}{m_Z}} \, .
\end{equation}
From the $WW$-threshold scan, an uncertainty of order $\Delta m_W/m_W \sim 4 \times 10^{-6}$ is achievable in the optimistic theory uncertainty scenario of~\cite{FCC:2025lpp} including parametric uncertainties, so a particle coupled via the Higgs portal shifts $m_W$ at a level that is expected to be measurable.\footnote{In the actual calculation of this effect we employ the evolution matrix formalism \cite{Fuentes-Martin:2020zaz} to solve the RGEs.}

Two caveats limit the use we make of this observable. Firstly, the one-loop matching of a heavy vector field is typically ill-defined in the absence of a more complete UV theory (though possible in principle~\cite{Thomsen:2024abg}), so we do not estimate the corresponding shift for the vector portal. Secondly, and more importantly, in the EFT, $C_{H\square}$ is directly related to the universal Higgs coupling modification $\kappaZZH$, so that $\Delta m_W$ largely repackages information already carried by the less precisely measured Higgs couplings rather than providing a theoretically independent constraint. We therefore include $m_W$ in the correlations shown in \cref{sec:discrimination} for illustration, motivated by~\cite{Maura:2024zxz}, where it was identified as a precise probe of the scalar portal, but do not include it in the quantitative discrimination analysis.

%%%%%%%%%%%%%%%%%%%%%%%%%%%%%%%%%%%%%%%%%%%%%%%%%%%%%
\section{Portal discrimination}
\label{sec:discrimination}
%%%%%%%%%%%%%%%%%%%%%%%%%%%%%%%%%%%%%%%%%%%%%%%%%%%%%
We now turn to the central question of this work: given a measured deviation in Higgs observables, can the underlying portal interaction be identified? We treat the on-shell and off-shell regimes separately. Below the invisible decay threshold, \cref{sec:discrimination-light} shows that the constraint from $\mathcal{B}_{\mathrm{inv}}$ leaves no observable handle to distinguish the portals, so that discrimination is only possible in the off-shell regime.
Focusing on this case, \cref{sec:discrimination-heavy} combines the observables of \cref{sec:pheno}, namely the direct WBF search and the indirect coupling and mass shifts $\Delta\kappaZZH, \Delta\kappaHHH, \Delta m_W$, into a single statistical framework. Since these indirect effects enter through one-loop contributions that depend on the spin and mass of the hidden state, the three portals trace out distinct trajectories in the space of observables even where any single measurement is degenerate between them. We use this to assess, for representative benchmarks, how much of the parameter space can be discovered and eventually also whether the portal responsible can be identified.

%%%%%%%%%%%%%%%%%%%%%%%%%%%%%%%%%%%%%%%%%%%%%%%%%%%%%
\subsection{The Higgs portal below threshold}
\label{sec:discrimination-light}
%%%%%%%%%%%%%%%%%%%%%%%%%%%%%%%%%%%%%%%%%%%%%%%%%%%%%
Decays into new final states---and in particular into two-particle final states---generally offer a uniquely sensitive probe of new physics. As discussed in \cref{sec:decays}, if decays into singlet final states are kinematically accessible, measurements of the Higgs invisible decay width are sensitive to an effective portal coupling down to $g_\text{eff}^2 \sim \brinv \times \Gamma_h / m_h \approx \brinv \times \num{e-4}$. 

Unfortunately, for higher-dimensional Higgs portal operators compatible with current invisible branching ratio constraints, \cref{eq:invBR} implies that the scale of the higher-dimensional Higgs portal operator is in the tens of TeV range in the case of a Higgs portal to fermions for an $\mathcal{O}(1)$ Wilson coefficient. The Higgs portal to vectors, on the other hand, could be in the TeV range or lower and hence potentially resolvable by direct searches for the mediators of the interaction. For Higgs portals below threshold, direct access to the mediator would be the only way of distinguishing portals since we find that any potential phenomenological effects due to the portal itself are either dominated by on-shell Higgs decays or too small to be seen given the invisible decay bounds. 

These other indirect effects imprinted by such weakly coupled Higgs portals are parametrically suppressed by roughly a loop factor compared to analogous SM corrections, and are therefore less constraining than the direct decay channel. The universal coupling shifts $\Delta\kappaZZH$ and the trilinear coupling modification $\Delta\kappaHHH$, discussed further below for the Higgs portal above threshold, cannot compete with the direct bound even at the precision goals of future colliders. We have checked that energy-growing effects expected from the higher-dimensional operators are also relatively negligible. This can be understood from the squared amplitude for any process involving a propagating Higgs boson producing two singlets in the final state, which scales with the momentum $p$ at most as
\begin{equation}
    \abs{\mathcal{M}}^2 \sim \frac{(p^2/\Lambda^2)^{2(d-4)}}{(p^2-m_h^2)^2 + m_h^2 \Gamma_h^2}
\end{equation}
for a Higgs portal of mass dimension $d$. Comparing this scaling for on-shell Higgs decays and the high-energy limit,
\begin{equation}
    \abs{\mathcal{M}}^2_\text{on-shell} \sim \frac{m_h^2}{\Gamma_h^2}\frac{m_h^{4(d-5)}}{\Lambda^{4(d-4)}} \, , \qquad
    \abs{\mathcal{M}}^2_\text{large-$p$} \sim \frac{p^{4(d-5)}}{\Lambda^{4(d-4)}} \, ,
\end{equation}
shows that energy-growing effects are significant only for $d>5$, and even then only for momenta far above the electroweak scale, $p^2 \sim (m_h/\Gamma_h)^{1/(d-5)} m_h^2$. The dimension-five fermion portal, in particular, gains nothing from this effect.

We therefore conclude that, within the observables discussed here and the currently proposed landscape of future colliders, a signature of an invisible Higgs decay could not be traced to a specific portal through effects due to the portal operator alone. For the vector portal, however, depending on the underlying UV completion, the heavy mediator between the Higgs and the new dark vector fields may be accessible to direct searches.

%%%%%%%%%%%%%%%%%%%%%%%%%%%%%%%%%%%%%%%%%%%%%%%%%%%%%%%%%%%%%%%%%%%%%%%%%%%%%%%%
\subsection{The Higgs portal above threshold}
\label{sec:discrimination-heavy}
%%%%%%%%%%%%%%%%%%%%%%%%%%%%%%%%%%%%%%%%%%%%%%%%%%%%%%%%%%%%%%%%%%%%%%%%%%%%%%%%
Higgs portals with singlets heavier than about $\SI{62.5}{GeV}$ do not contribute to the physical (on-shell) Higgs decay width and are searched for through a variety of other observable processes discussed in \cref{sec:pheno}, in particular Higgs coupling modifications in associated $Zh$ production and di-Higgs production, electroweak precision observables, and missing energy searches. In the following, we discuss if and how the broad experimental programme at future colliders can be leveraged to tell apart different scenarios of new physics with very similar, and particularly minimal, phenomenology.

%%%%%%%%%%%%%%%%%%%%%%%%%%%%%%%%%%%%%%%%%%%%%%%%%%%%%%%%%%%%%%%%%%%%%
\subsubsection{Kinematic distributions in weak boson fusion}
\label{sec:kinematic_distributions}
%%%%%%%%%%%%%%%%%%%%%%%%%%%%%%%%%%%%%%%%%%%%%%%%%%%%%%%%%%%%%%%%%%%%%

%%%%%%%%%%%%%%%%%%%%%%%%%%%%%%%%%%%%%%%%%%%%%%%%%%%%%%%%%%%%%%%%%%%%%
\begin{figure}
    \centering
    \includegraphics[width=\linewidth]{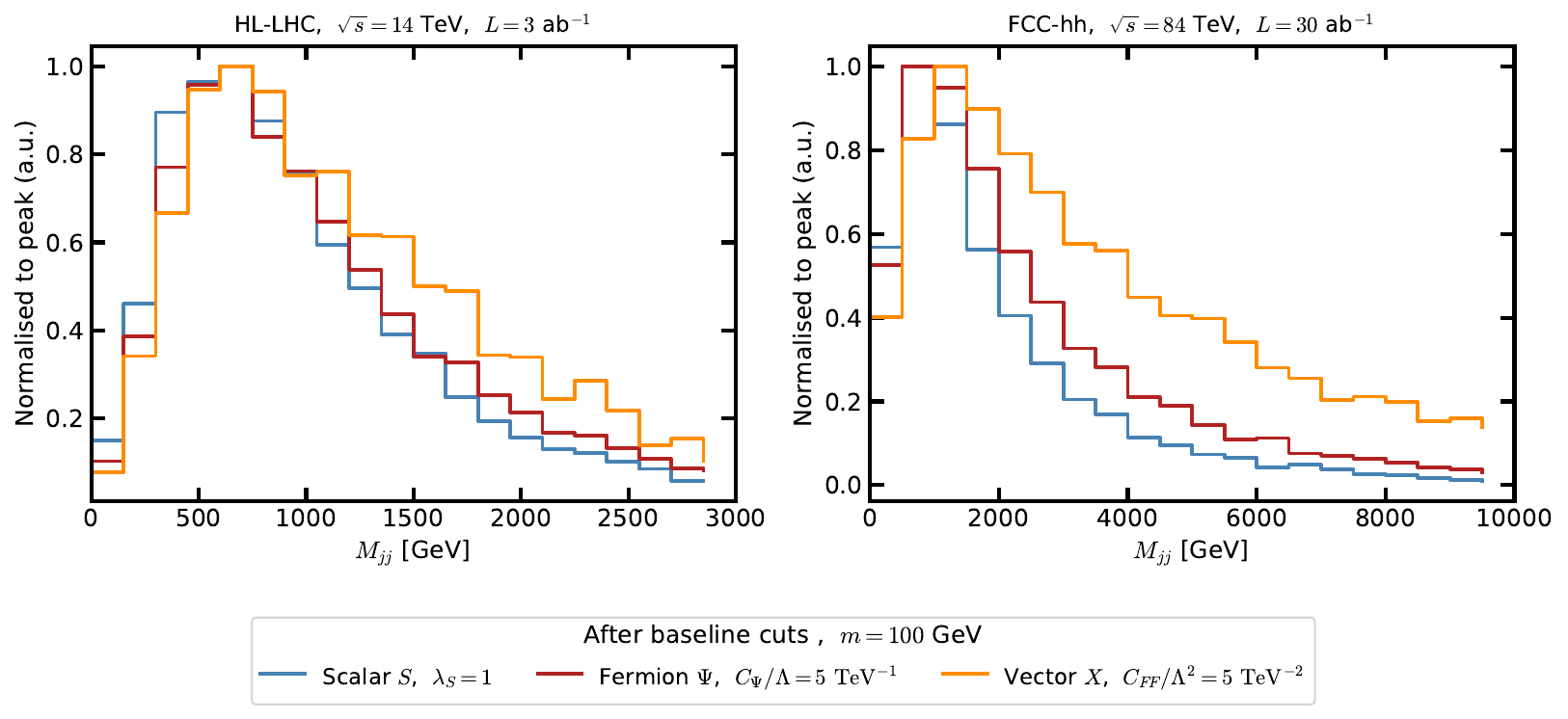}
    \caption{Normalised invariant mass of two jets in the weak boson fusion production of two invisible dark-sector particles through the scalar (blue), fermionic (red), or vector (orange) Higgs portals, for representative coupling values yielding a similar rate at HL-LHC and a light dark-sector mass of $M=\SI{100}{\giga\electronvolt}$.}
    \label{fig:mjj_lhc_vs_fcchh}
\end{figure}
%%%%%%%%%%%%%%%%%%%%%%%%%%%%%%%%%%%%%%%%%%%%%%%%%%%%%%%%%%%%%%%%%%%%%%
Higgs production in weak boson fusion is the most sensitive direct probe of the off-shell Higgs portal above threshold. A promising property to distinguish the portals is the mass dimension of the effective portal interactions, which leads to a relative energy enhancement $\sigma \sim ({E_{\text{CM}}/\Lambda})^{2(d-4)}$. One might, therefore, expect the kinematic distributions of the WBF production to contain enough information to disentangle the nature of a hypothetical signal. 

At the HL-LHC, the left panel of \cref{fig:mjj_lhc_vs_fcchh} shows the invariant mass distribution of the two jets in the weak boson fusion process consisting of two jets and missing transverse energy in the final state. The marginal Higgs portal operator to scalars is denoted by the histogram with a solid blue line, while the dimension-5 and dimension-6 Higgs portal operators to fermions and vectors are represented by histograms with solid red and orange lines, respectively. We see that while the higher-dimensional portals lead to slightly harder jets, this effect is mild, so fully disentangling the signal will require more information than contained in this single channel at HL-LHC alone. However, the right panel of \cref{fig:mjj_lhc_vs_fcchh} shows that benchmark points which are discoverable in WBF at HL-LHC can actually lead to distinguishable kinematic distributions at FCC-hh. 

Since the inclusive rates already capture the energy growth between HL-LHC and FCC-hh, we will focus next on the information contained in the rate only in evaluating the discriminating power of future colliders, though we note that a more thorough study of kinematical distributions in WBF at FCC-hh would further help disentangle the phenomenological profile of the Higgs portal operators.

%%%%%%%%%%%%%%%%%%%%%%%%%%%%%%%%%%%%%%%%%%%%%%%%%%%%%%%%%%%%%%%%%%%%%
\subsubsection{Model-dependent correlations among observables}
%%%%%%%%%%%%%%%%%%%%%%%%%%%%%%%%%%%%%%%%%%%%%%%%%%%%%%%%%%%%%%%%%%%%%
The different Higgs portals are generally capable of inducing the same departure from the SM prediction in a particular observable. However, the combination of multiple observables can help to differentiate between alternative scenarios of new physics if their dependence on the model and experimental parameters is sufficiently distinct.

Linear Higgs coupling modifications are controlled by the portals' radiative corrections to the (on-shell) Higgs two-point function, while the $hhh$ vertex also receives genuinely different one-particle irreducible (1PI) contributions, and direct effects leverage the distinct momentum dependence induced by the effective operators, implying a unique interplay between model and experimental parameters. To illustrate the correlations in observables for the different portals, we map the viable part of each portal's parameter space onto the plane spanned by two observables.
% 
%%%%%%%%%%%%%%%%%%%%%%%%%%%%%%%%%%%%%%%%%%%%%%%%%%%%%
\begin{figure}[t]
    \centering
    \includegraphics[scale=0.8]{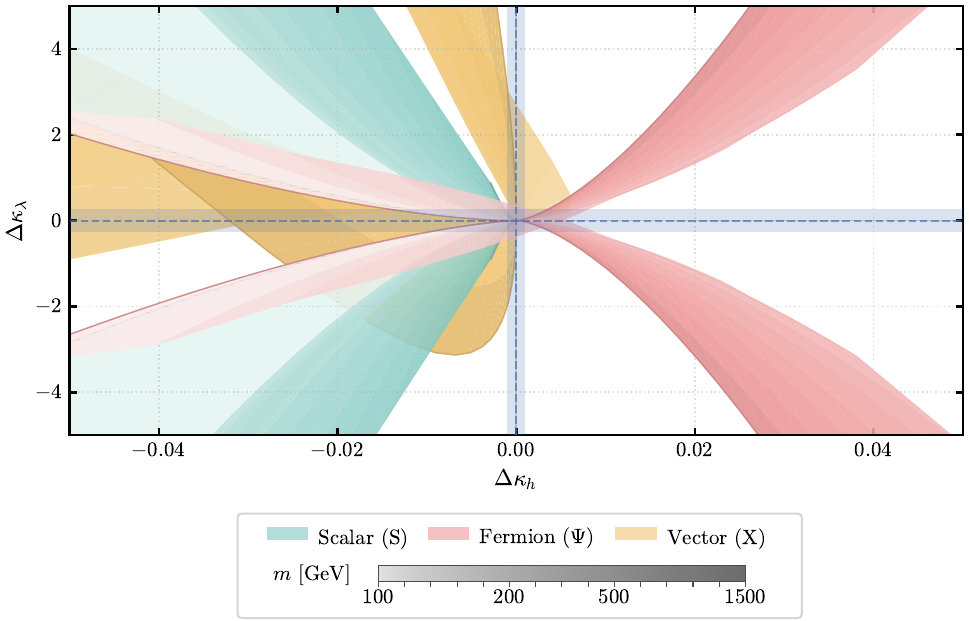}
    \caption{Parametric plot showing the trajectories in the space spanned by $(\Delta\kappaZZH,\Delta\kappa_{\lambda})$ as a function of the corresponding portal coupling for discrete values of the masses $m = \SIrange{100}{1500}{\giga\electronvolt}$, with lighter shaded slices corresponding to lighter masses and couplings $|\lambda_S|\leq 10$, $C_\psi/\Lambda \leq \SI{10}{\per\TeV}$ and $|\cff/\Lambda^2|\leq 10\,\si{\tera\electronvolt^{-2}}$. The blue bands indicate the projected sensitivity to $\Delta\kappaHHH = 0.275$ after HL-LHC and $\Delta\kappaZZH = 10^{-3}$ after FCC-ee (see \cref{tab:obs-summary}).}
    \label{fig:kappa_vs_kappa}
\end{figure}
%%%%%%%%%%%%%%%%%%%%%%%%%%%%%%%%%%%%%%%%%%%%%%%%%%%%%

As an example, we highlight the correlations in $\Delta \kappaZZH$ and the Higgs self-coupling $\Delta\kappaHHH$ in \cref{fig:kappa_vs_kappa}. The green, red and yellow bands indicate the trajectories in this parameter space as the corresponding portal couplings are varied for the scalar, fermion and vector portals, respectively. The mass of the dark sector particle is varied between 100 and 1500 GeV, corresponding to the shading going from lighter to darker. As $\kappaHHH$ is sensitive to up to three portal coupling insertions, the sign of the coupling becomes relevant for strong enough coupling, which is important for our analysis. This is in contrast with $\kappaZZH$: the wavefunction renormalisation related to the universal Higgs coupling modifier is proportional to the squared portal coupling (the linear contribution to the Higgs two-point function on the top-right of \cref{fig:feynman-diagrams} is absorbed in the Higgs boson mass renormalisation, see \cref{sec:renormalisation}). With these remarks in mind, we can analyse the coupling pattern correlations of the different portal interactions in more detail.

We start with the marginal scalar portal case. Here we observe $\Delta\kappaZZH\Delta\kappaHHH\propto -{\rm{sgn}}(\lambda_S)$. Therefore, the trajectories in the $\Delta\kappaZZH-\Delta \kappaHHH$ plane flow outwards irrespective of the scalar mass, approaching zero in the decoupling limit as $M\to\infty$. Considering the fermions as the first effective model, they are similarly narrowly localised in observable space for large masses, although they live in the opposite quadrant compared to the scalar. There is a small mass window where there is overlap with the scalar band, where the trajectory goes from $-+$ to $+-$ for positive couplings and $--\to++$ for negative couplings. However, sensitivity drops fast in that region. They also diverge as $M\to\infty$, owing to the effective nature of the model that remains sensitive to the UV scale~(see also~\cite{Brivio:2014pfa}). The vector portal displays a more involved $\kappaZZH-\kappaHHH$ correlation. Firstly, $\kappaZZH$ is predominantly negative as a function of the coupling and mass, except for a thin strip for masses $\sim 100\,-\,500 \si{\giga\electronvolt}$, overlapping with the small mass and coupling region of the fermions. The Higgs self-coupling modifier $\Delta\kappaHHH$, on the other hand, changes sign multiple times for a positive coupling, while it largely remains positive for negative coupling. Similarly, for large masses and small couplings, $\Delta\kappaHHH$ is typically positive, although this region blends with a negative $\kappaHHH$ region for large couplings and small mass. In the interpolating regime, for small couplings and large masses, $\Delta\kappaHHH$ can be positive, leading to the slightly awkward trace shown in \cref{fig:kappa_vs_kappa}.

Nevertheless, the following phenomenological picture emerges. Firstly, bosons can be distinguished from fermions via $\Delta \kappaHHH$ and $\Delta\kappaZZH$ in the large mass limit. Scalars and vectors are more difficult to distinguish given the correlations of the vector states since large positive couplings in the high-mass scalar region match onto small negative couplings in the large-mass region of the vectorial scenario. However, overall we see that there is only overlap in very specific regions of the parameter space for each pair of portals, enabling the possibility of at least partially discriminating between different models with these measurements alone. To fully disentangle any pair, more direct sensitivity through WBF will be necessary.

%%%%%%%%%%%%%%%%%%%%%%%%%%%%%%%%%%%%%%%%%%%%%%%%%%%%%
\begin{figure}[t]
    \centering
    \includegraphics[scale=0.9]{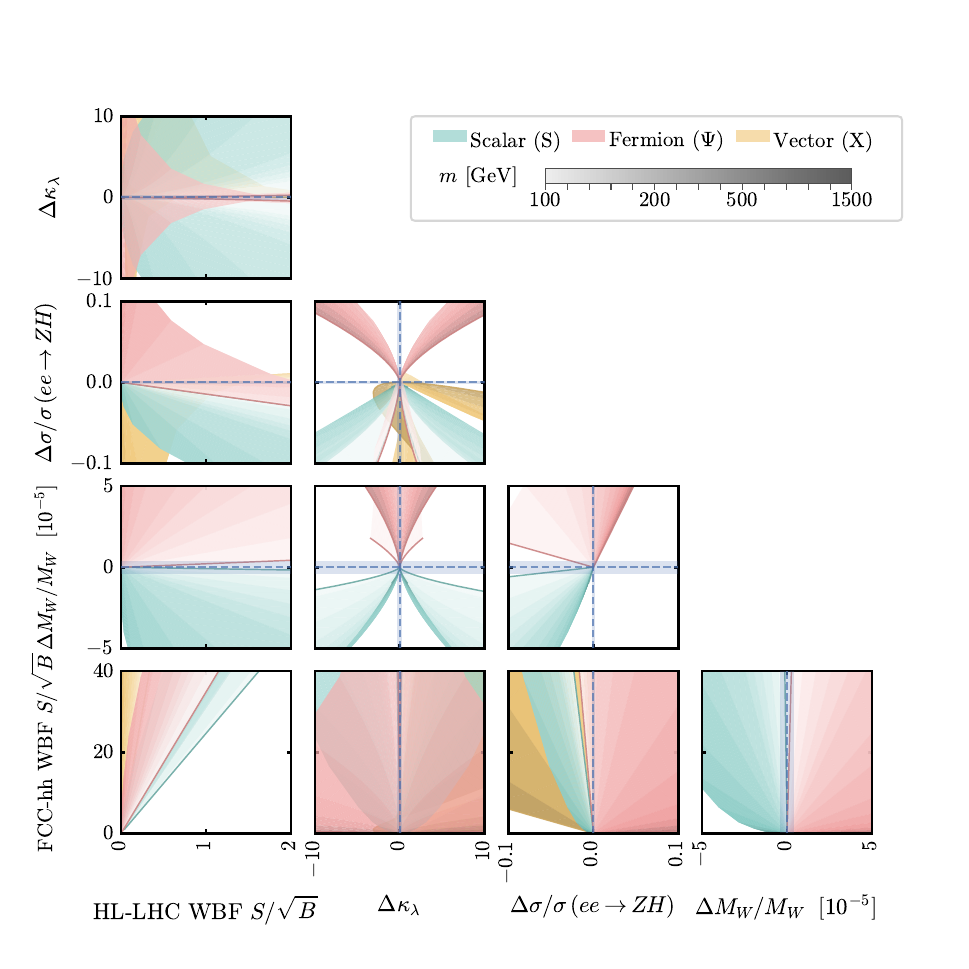}
    \caption{Same as \cref{fig:kappa_vs_kappa}, but for all observables included in our analysis. The shifts in the $W$ mass have been obtained by one-loop matching and running, which are omitted for the vector case. For the WBF observables, we show $S/\sqrt{B}$ as a proxy for the significance.}
    \label{fig:observable_correlations}
\end{figure}
%%%%%%%%%%%%%%%%%%%%%%%%%%%%%%%%%%%%%%%%%%%%%%%%%%%%%

The correlations among all observables discussed in \cref{sec:pheno} are presented in \cref{fig:observable_correlations}. Here, instead of the pseudo-observable $\kappaZZH$, we now use the relative deviation in the Higgsstrahlung process measurable at FCC-ee. Our qualitative assessment of \cref{fig:kappa_vs_kappa} extends to the collider processes shown; it is generally easier to separate bosons from fermions, but the bosons overlap in their phenomenology. Here, WBF can add sensitivity at hadron machines such as the HL-LHC or FCC-hh due to the higher-dimensional nature of the vector portal, which gives rise to a different missing-energy distribution that we exploit in our analysis through the modified total rate. The information in the relative rates for WBF at FCC-hh compared to HL-LHC is demonstrated in the bottom left plot of \cref{fig:observable_correlations}, where we see a different scaling in $S/\sqrt{B}$ for the different operator dimensions.  

We also show the shift in the $W$ mass from its SM value in \cref{fig:observable_correlations}, with the computation of this shift approximated by a one-loop matching and one-loop running. It is clear in all plots involving the $W$ mass that the scalar and fermion cases are shifted by opposite signs (we omit the vector case as one-loop matching in this case is ill-defined without further model-dependent considerations beyond our effective approach). This demonstrates the power that a detailed electroweak precision programme can add to the new physics programme at FCC-ee in general, and for the considered portal scenarios in particular. As we are limiting ourselves to Higgs-based observables in this paper and have not computed the full 2-loop contribution, we leave a more detailed exploration of this avenue for future work.

For the Higgs observables and for the portal boson vs. fermion discrimination, indirect effects are crucial. Since these are loop effects, the main difference comes from diverging signs. Only in the low-mass region where the deviation is small do we observe phenomenological degeneracy. But in this region, the double ratio of FCC-hh/HL-LHC WBF discriminates between different models very well. Put differently, indirect effects differ between bosons and fermions in the large mass regime and diverge in sign. They can only mimic each other by matching a large-mass point for a vector with a small-mass point for the fermion, for which the WBF effects are too large, so bosons and fermions are distinguishable (except for small and isolated parameter regions). In this sense, present and future WBF measurements are critical ingredients for portal spectroscopy once a discovery is made.

%%%%%%%%%%%%%%%%%%%%%%%%%%%%%%%%%%%%%%%%%%%%%%%%%%%%%%%%%%%%%%%%%%%%%
\subsubsection{Criterion for distinguishability}
%%%%%%%%%%%%%%%%%%%%%%%%%%%%%%%%%%%%%%%%%%%%%%%%%%%%%%%%%%%%%%%%%%%%%

%%%%%%%%%%%%%%%%%%%%%%%%%%%%%%%%%%%%%%%%%%%%%%%%%%%%%
\begin{figure}[!t]
    \centering
    \includegraphics[width=\linewidth]{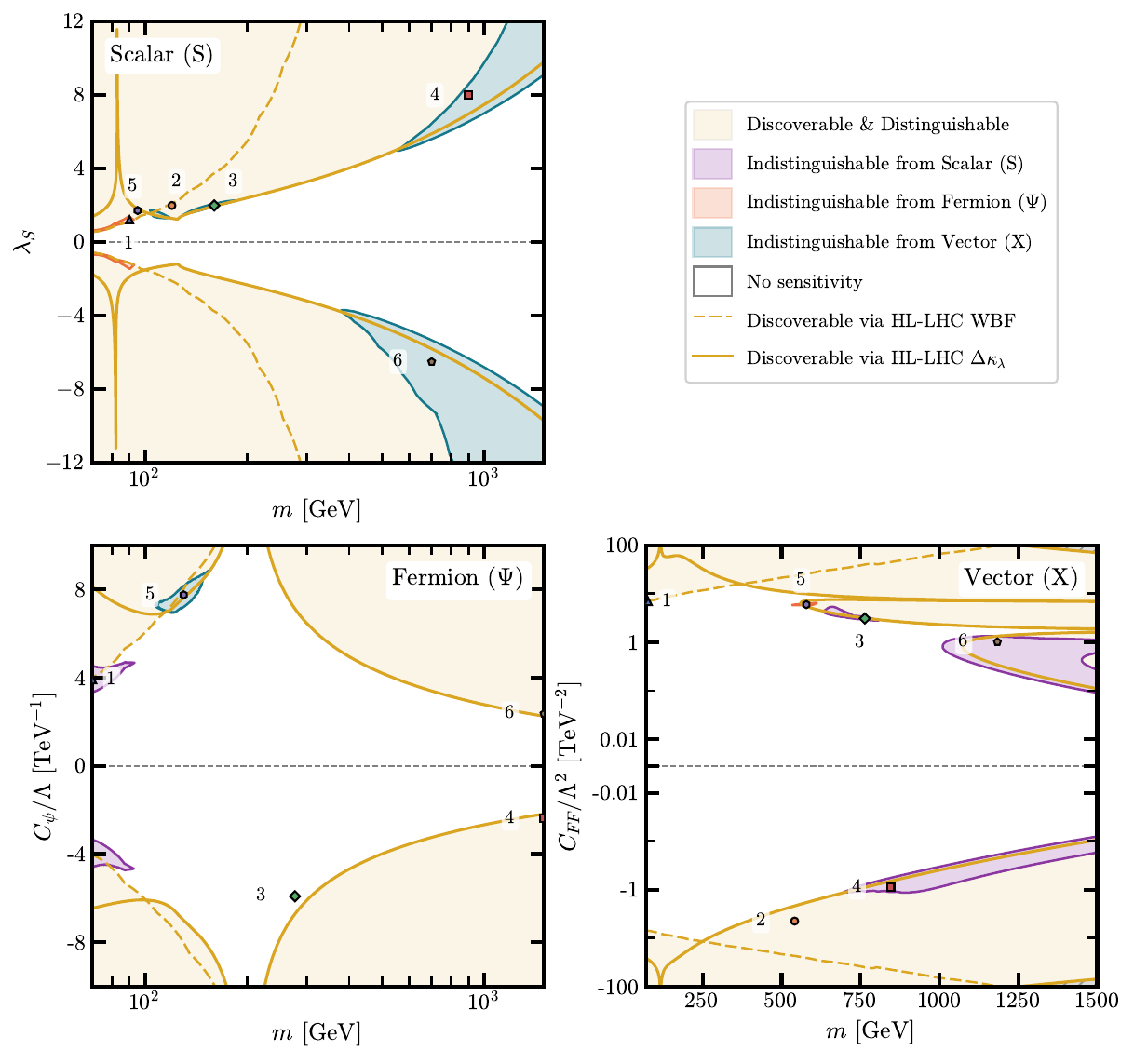}
    \caption{Regions of parameter space which are discoverable at $3\sigma$ by either WBF (dashed yellow line) or di-Higgs production (solid yellow line) at HL-LHC. Purple (green) regions cannot be distinguished from the scalar (vector) case at $3\sigma$ even after FCC-hh. The orange region around benchmark points 1 and 5 are barely visible, corresponding to indistinguishability from the fermion case. We include all observables except for $m_W$ in assessing distinguishability.}
    \label{fig:confusion_discovery_hllhc}
\end{figure}
%%%%%%%%%%%%%%%%%%%%%%%%%%%%%%%%%%%%%%%%%%%%%%%%%%%%%

Since the trajectories through parameter space differ significantly across the portal scenarios, combining all observables could allow us to disentangle them to a reasonable degree. To quantify this statement, we define a likelihood assuming a hypothetical measurement $(\mathcal{O}_k^{\rm exp},N_k^{\rm exp})$ as: 
\begin{equation}
    \label{eq:likelihood}
    -2\ln\mathcal{L}(p_A)
    = \sum_{k\,\in\,\mathrm{Gauss}} \frac{\big[\mathcal{O}_k(p_A)-\mathcal{O}_{k}^{\rm exp}\big]^2}{\sigma_k^2}
    + \sum_{k\,\in\,\mathrm{Pois}} 2\left[\mu_k(p_A)-N_{k}^{\rm exp} + N_{k}^{\rm exp}
      \,\ln\frac{N_{k}^{\rm exp}}{\mu_k(p_A)}\right],
\end{equation} 
with $p_A$ a point in the parameter space of model A and $\mu_k = S_k + B_k$ the expected WBF yield. We assume WBF production follows a Poisson distribution, while all other measurements are Gaussian. 

We can now define a criterion for whether a given channel registers a large enough deviation from the SM, which we refer to as \emph{discoverability}; and second, a measure of whether two models remain experimentally indistinguishable once the available channels are taken into account, which we refer to as \emph{indistinguishability}. For {\bf discoverability}, we look at the predictions generated by a parameter point $p_A$ of model $A$ in an individual channel and compare it to the SM. We only consider points with perturbative couplings and for which the EFT expansion is well-defined by requiring $|c| \leq \frac{(4\pi)^{n_c}}{m^{\,d-4}}.$ with $n_c=1(2)$ for the marginal (higher-dimensional) portal. The resulting log-likelihood then follows a one-dimensional $\chi^2$ distribution, and discovery at $n_\sigma$ significance can be defined as $-2\ln\mathcal{L}(p_A) = (n_\sigma )^2$. 

We now describe our procedure to distinguish from model $A$ an alternative hypothesis provided by a point $q_B$ in the parameter space of model B. Since we are assuming the $n_\sigma$ discovery of a signal, only points $q_B$ leading to the same discovery threshold can mimic $p_A$. Requiring exact equality is too strong a criterion, so instead we consider all points which reproduce the same signal to $\pm 1 \sigma$. This defines a region of interest for which we can define the notion of {\bf indistinguishability} $\mathcal I_{A\to B}(p_A,q_B)$. We fix the predictions from model $A$ as the null hypothesis by setting $(O_k^{\rm exp}, N_k^{\rm exp}) = (O_k(p_A), N_k(p_A))$ in \cref{eq:likelihood} to define the following test statistic 
\begin{equation}
\label{eq:lrt-q}
q_{A\to B}(p)
= \min_{q\in\mathcal{P}_B} \big[-2\ln\mathcal{L}(q,p)\big],
\end{equation}
which is derived from the likelihood ratio between the two models. Naively, one could simply use Wilks' theorem~\cite{Wilks:1938dza} here and assume $q_{A\to B}$ to follow an $n$-dimensional $\chi^2$ distribution. However, it does not apply here since $A$ and $B$ are non-nested overlapping models, so to define the criterion of indistinguishability more rigorously one would need to calibrate the distribution of $q$ through toy measurements. It is sufficient for our illustrative purposes here to simply define our threshold criteria assuming the distribution indeed approaches an $n$-dimensional $\chi^2$ statistic, where $n = n_{\rm obs} - 2$ as we are profiling over two parameters. Thus, two points are indistinguishable at $n_\sigma$ if 
\begin{equation}
q_{A\to B}(p_A)
<
\chi^2_{n_{\rm obs}-2;\,1-\alpha},  \qquad \alpha =  2\big[1-\Phi(n_\sigma)\big]. 
\end{equation}

The resulting discoverable and indistinguishable regions of parameter space in the mass vs coupling plane for a given portal interaction at $\geq3\sigma$ are shown in \cref{fig:confusion_discovery_hllhc}, with the marginal portal to scalars on the top left, the dimension-5 portal to fermions on the bottom left, and the dimension-6 portal to vectors on the bottom right. The yellow contours delimited by a dashed and solid line correspond, respectively, to points discoverable via WBF or di-Higgs production at HL-LHC. We see that the potential discoverability region of di-Higgs measurements extends out to larger masses than for WBF, though we note that the computation for the effective operators becomes ambiguous at large masses, as discussed in \cref{sec:oneloopconvergence}. The phenomenology of the alternative hypothesis is indistinguishable from that of a scalar portal in the purple-shaded region, from the fermion portal in the orange-shaded region, and from the vector portal in the green-shaded regions; there are no regions for which the base hypothesis is indistinguishable from both alternative portals. In other words, it is always possible to distinguish at least one case from the other two, and all three scenarios can be distinguished from each other in the yellow-shaded region that does not overlap with the purple, orange or green regions. 

%%%%%%%%%%%%%%%%%%%%%%%%%%%%%%%%%%%%%%%%%%%%%%%%%%%%%
\begin{table}[!t]
\centering
\caption{Breakdown of the predictions from each model for the point that best mimics the signal for the benchmark points displayed in \cref{fig:confusion_discovery_hllhc}. The model that constitutes the null hypothesis is denoted by a 0 in the last column and the discovery channel at HL-LHC is reproduced within $1\sigma$. The colour coding indicates whether a portal is distinguishable (blue) from the benchmark, or confusable with the fermion (red) or the vector (yellow) portal.}
\vspace{0.25cm}
\resizebox{\textwidth}{!}{%
\begin{tabular}{lcccccccc}
\hline\hline
\multicolumn{1}{c|}{\multirow{2}{*}{Model}} &
\multicolumn{1}{c|}{\multirow{2}{*}{Coupling}} &
\multicolumn{1}{c|}{\multirow{1}{*}{Mass}} &
\multicolumn{1}{c|}{\multirow{1}{*}{Discov. at}} &
\multicolumn{1}{c|}{HL-LHC} &
\multicolumn{1}{c|}{FCC-ee $Zh$} &
\multicolumn{2}{c|}{FCC-hh} &
\multicolumn{1}{c}{\multirow{1}{*}{Indisting.}} \\
\cline{7-8}
\multicolumn{1}{c|}{} &
\multicolumn{1}{c|}{} &
\multicolumn{1}{c|}{[GeV]} &
\multicolumn{1}{c|}{HL-LHC} &
\multicolumn{1}{c|}{WBF $S/\sqrt{B}$} &
\multicolumn{1}{c|}{$\Delta\sigma/\sigma$ [$10^{-2}$]} &
\multicolumn{1}{c|}{WBF $S/\sqrt{B}$} &
\multicolumn{1}{c|}{$\Delta\kappaHHH$} &
\multicolumn{1}{c}{at $n_\sigma$} \\
\hline\hline

\multicolumn{9}{l}{\textbf{Benchmark 1}} \\

\rowcolor{confFermion!40} S & $\hphantom{-}1.23$ & 90 & WBF & $3.5$ & $-3.97 \pm 0.21$ & $76.1$ & $\hphantom{-}0.12 \pm 0.05$ & $0$ \\

\rowcolor{confFermion!40} $\psi$ & $\hphantom{-}3.9\,\si{\tera\electronvolt^{-1}}$ & 70 & WBF & $2.8$ & $-3.53 \pm 0.21$ & $77.2$ & $\hphantom{-}0.05 \pm 0.05$ & $2.0\sigma$ \\

\rowcolor{distinguishable!40} X & $\hphantom{-}6.9\,\si{\tera\electronvolt^{-2}}$ & 77 & WBF & $2.8$ & $-0.32 \pm 0.21$ & $303$ & $-0.00 \pm 0.05$ & $222\sigma$ \\

\midrule

\multicolumn{9}{l}{\textbf{Benchmark 2}} \\

\rowcolor{distinguishable!40} S & $\hphantom{-}2$ & 120 & $\Delta\kappaHHH$ & $2.7$ & $-4.58 \pm 0.21$ & $70.9$ & $\hphantom{-}3.13 \pm 0.05$ & $0$ \\

\rowcolor{distinguishable!40} $\psi$ & $-10.1\,\si{\tera\electronvolt^{-1}}$ & 275 & $\Delta\kappaHHH$ & $0.8$ & $\hphantom{-}6.43 \pm 0.21$ & $72.6$ & $\hphantom{-}2.71 \pm 0.05$ & $53.1\sigma$ \\

\rowcolor{distinguishable!40} X & $-4.4\,\si{\tera\electronvolt^{-2}}$ & 541 & $\Delta\kappaHHH$ & $0.1$ & $-1.66 \pm 0.21$ & $70.8$ & $\hphantom{-}3.21 \pm 0.05$ & $13.9\sigma$ \\

\midrule

\multicolumn{9}{l}{\textbf{Benchmark 3}} \\

\rowcolor{confVector!40} S & $\hphantom{-}2$ & 160 & $\Delta\kappaHHH$ & $0.9$ & $-2.29 \pm 0.21$ & $26.4$ & $\hphantom{-}0.87 \pm 0.05$ & $0$ \\

\rowcolor{distinguishable!40} $\psi$ & $-5.9\,\si{\tera\electronvolt^{-1}}$ & 277 & $\Delta\kappaHHH$ & $0.3$ & $\hphantom{-}2.22 \pm 0.21$ & $24.5$ & $\hphantom{-}0.57 \pm 0.05$ & $22.1\sigma$ \\

\rowcolor{confVector!40} X & $\hphantom{-}3.1\,\si{\tera\electronvolt^{-2}}$ & 764 & $\Delta\kappaHHH$ & $0$ & $-2.81 \pm 0.21$ & $26.6$ & $\hphantom{-}0.79 \pm 0.05$ & $2.4\sigma$ \\

\midrule

\multicolumn{9}{l}{\textbf{Benchmark 4}} \\

\rowcolor{confVector!40} S & $\hphantom{-}8$ & 900 & $\Delta\kappaHHH$ & $0$ & $-1.02 \pm 0.21$ & $1.13$ & $\hphantom{-}1.24 \pm 0.05$ & $0$ \\

\rowcolor{distinguishable!40} $\psi$ & $-2.4\,\si{\tera\electronvolt^{-1}}$ & 1500 & $\Delta\kappaHHH$ & $0$ & $\hphantom{-}1.84 \pm 0.21$ & $0.09$ & $\hphantom{-}1.06 \pm 0.05$ & $13.8\sigma$ \\

\rowcolor{confVector!40} X & $-0.9\,\si{\tera\electronvolt^{-2}}$ & 845 & $\Delta\kappaHHH$ & $0$ & $-0.33 \pm 0.21$ & $2.07$ & $\hphantom{-}1.25 \pm 0.05$ & $2.6\sigma$ \\

\midrule

\multicolumn{9}{l}{\textbf{Benchmark 5}} \\

\rowcolor{distinguishable!40} S & $\hphantom{-}1.73$ & 95 & $\Delta\kappaHHH$ & $5.0$ & $-6.60 \pm 0.21$ & $118$ & $\hphantom{-}0.70 \pm 0.05$ & $13.2\sigma$ \\

\rowcolor{confVector!40} $\psi$ & $\hphantom{-}7.8\,\si{\tera\electronvolt^{-1}}$ & 130 & $\Delta\kappaHHH$ & $2.8$ & $-3.94 \pm 0.21$ & $121$ & $\hphantom{-}0.87 \pm 0.05$ & $0$ \\

\rowcolor{confVector!40} X & $\hphantom{-}5.9\,\si{\tera\electronvolt^{-2}}$ & 579 & $\Delta\kappaHHH$ & $0.2$ & $-3.90 \pm 0.21$ & $121$ & $\hphantom{-}0.88 \pm 0.05$ & $1.9\sigma$ \\

\midrule

\multicolumn{9}{l}{\textbf{Benchmark 6}} \\

\rowcolor{confVector!40} S & $-6.5$ & 700 & $\Delta\kappaHHH$ & $0$ & $-1.11 \pm 0.21$ & $2$ & $-1.16 \pm 0.05$ & $0$ \\

\rowcolor{distinguishable!40} $\psi$ & $\hphantom{-}2.4\,\si{\tera\electronvolt^{-1}}$ & 1500 & $\Delta\kappaHHH$ & $0$ & $\hphantom{-}1.83 \pm 0.21$ & $0.09$ & $-0.96 \pm 0.05$ & $14.3\sigma$ \\

\rowcolor{confVector!40} X & $\hphantom{-}1.0\,\si{\tera\electronvolt^{-2}}$ & 1184 & $\Delta\kappaHHH$ & $0$ & $-1.13 \pm 0.21$ & $1.78$ & $-1.16 \pm 0.05$ & $0.0\sigma$ \\

\bottomrule
\end{tabular}
}
\label{tab:benchmarks}
\end{table}
%%%%%%%%%%%%%%%%%%%%%%%%%%%%%%%%%%%%%%%%%%%%%%%%%%%%%
%%%%%%%%%%%%%%%%%%%%%%%%%%%%%%%%%%%%%%%%%%%%%%%%%%%%%

%%%%%%%%%%%%%%%%%%%%%%%%%%%%%%%%%%%%%%%%%%%%%%%%%%%%%
\subsubsection{Can we disentangle points discoverable by WBF?}
%%%%%%%%%%%%%%%%%%%%%%%%%%%%%%%%%%%%%%%%%%%%%%%%%%%%%
Let us unravel this result by focusing on one discovery channel at a time, starting with points that can be reached through WBF at HL-LHC. A helpful guide in this exercise will be \cref{fig:kappa_dependence}, which shows the Higgs coupling and self-coupling modifications induced by the different portals, overlaid with the discoverable and indistinguishable regions from \cref{fig:confusion_discovery_hllhc} in the plane of mass vs couplings. The red and blue regions of \cref{fig:kappa_dependence} indicate positive and negative values of $\Delta\kappa_\lambda$ and $\Delta\kappa_h$ on the left and right plots, respectively, and the scalar, fermion, and vector cases correspond to the top, middle and bottom plots. The benchmark points (BP) detailed in \cref{tab:benchmarks}, which are highlighted on both figures, further help dissect the underlying physics. 

%%%%%%%%%%%%%%%%%%%%%%%%%%%%%%%%%%%%%%%%%%%%%%%%%%%%%
\newenvironment{ndescription}{%
   \begin{description}[
    style=sameline,
    leftmargin=0.4cm,
    font=\bfseries,
    itemsep=0.2em
  ]%
}{%
   \end{description}%
}
%%%%%%%%%%%%%%%%%%%%%%%%%%%%%%%%%%%%%%%%%%%%%%%%%%%%%

%%%%%%%%%%%%%%%%%%%%%%%%%%%%%%%%%%%%%%%%%%%%%%%%%%%%%
\begin{ndescription}
\item[Scalar vs Fermion.] The first thing to note is that reproducing the HL-LHC signal without triggering a deviation in di-Higgs production that would already have been observed forces both models into the restricted low-mass region, $m_S<\SI{150}{\giga\electronvolt}$ and $m_{\psi}<\SI{250}{\giga\electronvolt}$. This is because a sufficiently large WBF missing-energy signal requires comparatively large couplings. In this region of parameter space, the scalar and fermion portals can mimic one another in di-Higgs production through a rather specific correlation between masses and couplings, as can be extracted from \cref{fig:kappa_dependence}. This effectively renders the scenarios indistinguishable at the HL-LHC.

%%%%%%%%%%%%%%%%%%%%%%%%%%%%%%%%%%%%%%%%%%%%%%%%%%%%%
\begin{figure}[p]
    \centering
    %\hspace{-1cm}
    \includegraphics[width=\textwidth]{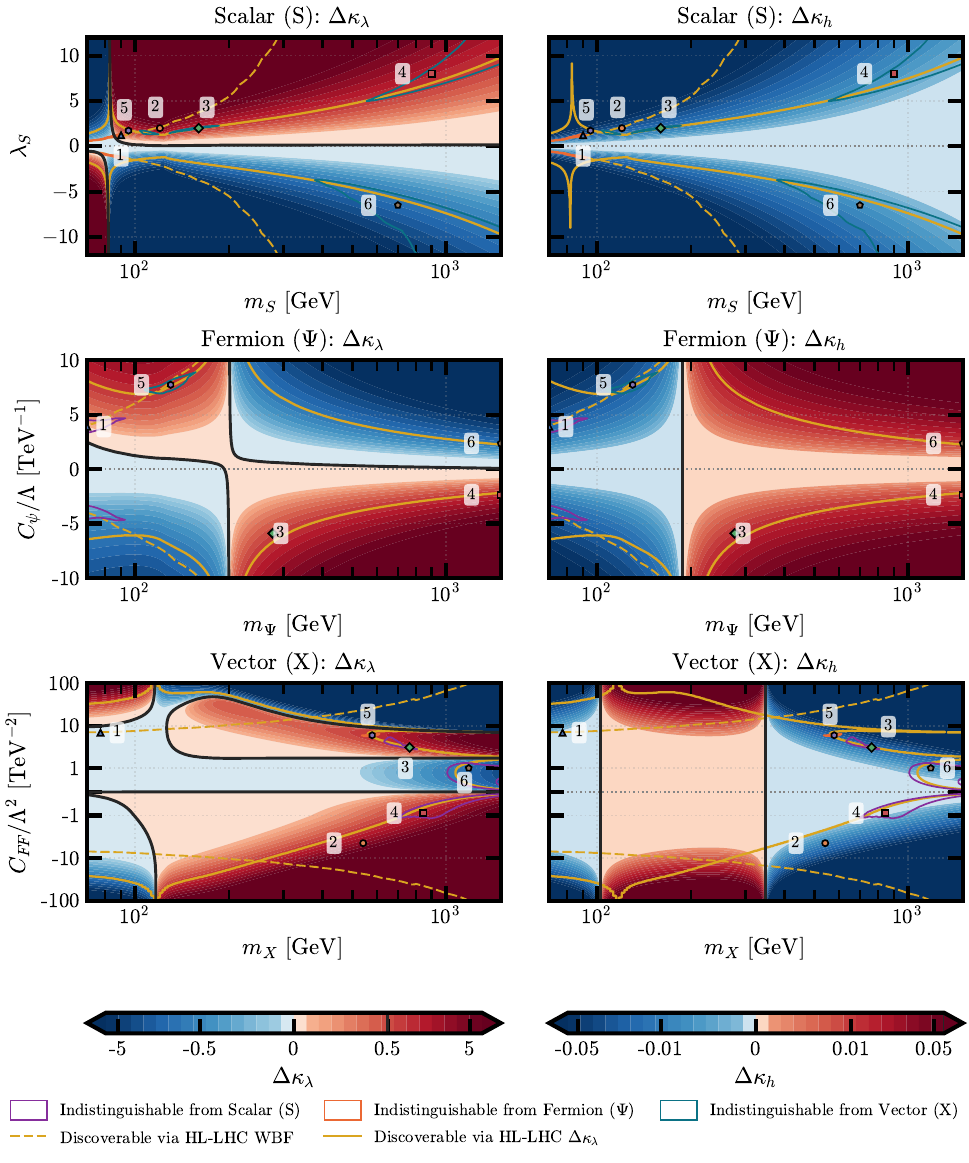}
    \caption{Contour plot in the plane of mass vs coupling showing $\Delta \kappaHHH$ (left) and $\Delta\kappaZZH$ (right) for the scalar, fermionic and vector portals from top to bottom. The points shown correspond to the benchmark points in \cref{tab:benchmarks}. The colour coding corresponds to a signed logarithmic scale centred around 0, with positive (negative) values in increasingly darker shades of red (blue). The contours along which the sign of the coupling modifiers changes are denoted by a solid black line. The discoverable and (in)distinguishable regions from \cref{fig:confusion_discovery_hllhc} are overlaid as unshaded contours following the same colour coding.}
    \label{fig:kappa_dependence}
\end{figure}
%%%%%%%%%%%%%%%%%%%%%%%%%%%%%%%%%%%%%%%%%%%%%%%%%%%%%

Including the information from Higgsstrahlung measurements at FCC-ee only reduces the parameter space towards even smaller masses, since this regime coincides precisely with the region for which the observable trajectories in \cref{fig:kappa_vs_kappa} overlap. The FCC-ee determination of $m_W$ could provide additional discrimination in this region, although this would require a dedicated two-loop calculation. Firstly, the corresponding effects are small for the low masses and couplings that survive the previous constraints. This conclusion may, however, depend partly on the way the shift is evaluated here, namely through NLO SMEFT running. A sizeable contribution from operator mixing requires a sufficiently large hierarchy of scales. In contrast, in our setup the UV scale is identified with the mediator mass, leaving little running and hence only a small mixing contribution into $C_{HD}$. A dedicated two-loop calculation need not exhibit the same suppression (e.g.\ \cite{Englert:2020gcp}). Additionally, the EFT validity is brought into question precisely in the region of interest. The behaviour of large threshold effects not correctly captured by the EFT at leading order could likewise affect this statement.  

The contribution from di-Higgs measurements at FCC-hh further enhances the ability to distinguish between the scenarios. The points which have passed the filters for $ee\to Zh$ and the HL-LHC depend on the freedom associated with $\kappaHHH$. When the precision achieved at FCC-hh goes beyond the projection of HL-LHC, a tension between the models begins to appear. The final discriminator is WBF production at FCC-hh. Having applied all the previous filters, the scalar portal is left with smaller masses and couplings than the fermion portal and thus has a systematically lower total rate. Together with the fact that the WBF signal shows a different energy dependence in the two cases, this is enough to distinguish almost all points along the remaining degenerate trajectories. 

Only a small island surrounding BP1 survives, a narrow region close to the discovery threshold where the scalar and fermion portals cannot be distinguished, surrounded by otherwise distinguishable parameter points. Even a modest variation of $\Delta\lambda_S\sim\mathcal{O}(0.1)$ is sufficient to remove it. Further evidence of the frailty of this region is the fact that these two hypotheses would still exhibit a $2\sigma$ tension, so there would at least be a preference for one scenario over the other in this event. 

%%%%%%%
\item[Fermion vs Vector.] Similar to the fermion vs scalar case, the HL-LHC alone cannot distinguish the two models as a single experiment. That said, it will significantly restrict the allowed mediator-mass range. Requiring, in addition, that the two scenarios reproduce the same signature at FCC-ee reduces the degeneracy to a small region of parameter space around $m_X\sim\SI{400}{\giga\electronvolt}$. In this region, WBF is sensitive, and the predicted rates differ between the two models to an extent that the remaining degeneracy can mostly be resolved.

Again, a small island very close to the discovery threshold survives surrounding BP5, which coincides with the small region in \cref{fig:kappa_dependence} showing an overlap, although achieving the same phenomenological signatures is only possible if one accepts a mild tension of $1.9\sigma$.
%%%%%%%
\item[Scalar vs Vector.]A similar picture emerges in this case: reproducing the same signatures until after the FCC-ee requires specific correlations between masses and couplings. This leaves only a limited degeneracy between the two portal scenarios. Resolving this remaining ambiguity becomes possible with direct FCC-hh measurements. 
\end{ndescription}
%%%%%%%%%%%%%%%%%%%%%%%%%%%%%%%%%%%%%%%%%%%%%%%%%%%%%

An interesting overall feature of the model comparison is that the points which are discoverable through WBF force the parameter space to small portal masses. There, distinguishability from indirect measurements can only restrict the confusable region of parameter space, but a reliable conclusion generally requires a direct, energy-enhanced measurement. It is the combination of both observations that enables us to disentangle the models from each other.

\subsubsection{Can we disentangle points discoverable by di-Higgs production?}
We now turn our attention to points discoverable through di-Higgs production. With it as our anchor, the non-trivial parametric dependence of the coupling modifiers, as well as the fact that the observable is now sensitive to the sign of the coupling, makes this situation a bit harder to follow. Firstly, all portals can mimic each other's signal for at least some region of parameter space for just this one measurement (see \cref{fig:kappa_dependence}), but as evidenced by \cref{fig:confusion_discovery_hllhc}, the combination of all observables retains enough distinguishing power to mostly disentangle the nature of the Higgs portal. 

\begin{ndescription}
%%%%%%%
\item[Scalar vs Fermion.] 
The situation here is rather simple. In the low mass regime, points discoverable by di-Higgs production are also discoverable via WBF, and we have already established that the two portals can be disentangled there. Above the threshold of masses accessible via WBF, to which we retain sensitivity for strong enough couplings, we need information from the FCC-ee. There, a dark sector coupled through the scalar portal would reduce the Higgsstrahlung cross-section, while one coupled through the fermion portal would increase it, which suffices to fully disentangle the two portals. 
%%%%%%%
%%%%%%%
\item[Fermion vs Vector.] 
%%%%%%%
The conclusion is similar to that for the scalar portal, with the exception of the small island surrounding BP5, which was already present when discussing WBF in the previous section, as both probes are similarly sensitive in that region.  
%%%%%%%
\item[Scalar vs Vector.]  
%%%%%%%
Just as heralded by the correlation plot in \cref{fig:kappa_vs_kappa}, this combination is the hardest to disentangle. Let us first focus on the high-mass region. Once HL-LHC sensitivity from WBF dies off, we need information from the FCC-ee to make progress, but in the large mass limit the vector and scalar portals lead to an identical imprint on the Higgsstrahlung cross-section, so this filter only tightens the allowed region of parameter space.
If FCC-hh is still sensitive enough, then the energy enhancement of the vector portal interaction suffices to disentangle the portals. For the scalar, WBF sensitivity tails off for masses $m_S\sim\SI{750}{\GeV}$, so detectable coupling-mass combinations above that essentially present a degenerate system with only two handles and two independent parameters. This renders the two portals indistinguishable in the regions surrounding BP4 and BP6.

One thing to note here is that for the scalar portal, $\Delta\kappaHHH$ goes to zero as a function of the mass, controlled by the new physics coupling. For the vector portal, we find $\lim_{M_X\to\infty} \Delta\kappaHHH = \pm\infty$, due to the EFT nature of the model, which imparts non-decoupling effects. The only reason the indistinguishable region just discussed exists is because a sufficiently strong coupling modifier due to the vector portal interaction, as required to mimic the scalar portal, can be found by increasing the mass and reducing the coupling, which simultaneously reduces the WBF at FCC-hh. This is further evidenced by the fact that in the vector plane, this region extends to smaller couplings and larger masses. Thus, the existence of this region relies on divergent contributions that grow with heavier masses, which brings into question the reliability of our results there (see \cref{sec:oneloopconvergence} for a more detailed discussion). This region is cut off by requiring the calculation to remain valid, but in the interim and without specifying a UV completion for this interaction, the possibility of confusing the scalar and vector portals remains open here.

Another interesting feature of these regions is that it is easier to fake a reduction in the di-Higgs signal than an increase. For example, for BP6 the vector portal can precisely fake the scalar portal, while it leads to a $2.6 \sigma$ tension for the corresponding BP4, which is a representative of the analogous region for positive $\lambda_S$. 

We now shift our attention to the low-mass region points not discoverable by WBF, which puts us at masses $m_S\sim \mathcal{O}(100-200\,\si{\giga\electronvolt})$ and $\lambda_S\sim\mathcal{O}(1)$. Similar to the WBF case, though, the requirement of a mild enough signal from WBF at FCC-hh and good enough reproduction of the Higgsstrahlung cross-section proves difficult enough to mimic, so only two small regions surrounding BP3 can actually achieve it, at the cost of a $\gtrsim 2\sigma$ tension. The sharp increase in the indistinguishability metric to $13.9\sigma$ for BP2 further indicates how frail the equilibrium found in this region is.
\end{ndescription}

To summarise, if the extended Higgs sector cannot be probed directly through on-shell Higgs-boson decays, only a limited number of phenomenological observables remain with which to uncover the portal. Nevertheless, our analysis has revealed that bosonic portal interactions can be distinguished from fermionic ones, thanks to the distinct high-mass behaviour of the loop-induced Higgs-coupling modifiers. Interactions with scalars and vectors are harder to tell apart based on this metric alone, so the distinguishing power crucially relies on the impact that the dimension of the portal has on WBF production at FCC-hh. The latter dies off at a certain mass threshold, after which a scalar and vector portal interaction can no longer be disentangled reliably.\footnote{We note that the renormalisation scale dependence discussed in \cref{sec:renormalisation} does not alter our qualitative conclusions; it shifts the overlap regions to different mass-coupling correlations, but as long as these remain within the parameter space accessible via WBF one is still able to disentangle them irrespective of the choice of renormalisation scale.}

Therefore, even if the reduced set of phenomenological signatures this scenario presents makes discrimination challenging, we conclude that once a discovery has been made the different scenarios can largely be distinguished through the broader, multifaceted Higgs phenomenology programme integrating different collider stages.

%%%%%%%%%%%%%%%%%%%%%%%%%%%%%%%%%%%%%%%%%%%%%%%%%%%%%
\section{Summary and conclusions}
\label{sec:conc}
%%%%%%%%%%%%%%%%%%%%%%%%%%%%%%%%%%%%%%%%%%%%%%%%%%%%%
Overwhelming cosmological and astrophysical data point to the existence of dark matter, which could very well be part of a larger hidden landscape of particle physics beyond the Standard Model. The Higgs field in the SM is a unique bridge between the SM and these possible dark sector interactions, with collider-relevant consequences. As measurements of the Higgs sector enter a new precision era with the HL-LHC around the corner and possible next-generation Higgs factories on the horizon, we may well see evidence for a Higgs portal appear in data.  

Whilst the standard, marginal Higgs portal interaction with dark sector scalars is theoretically motivated from the perspective of renormalisability, it is only one option for how the Higgs could interact with a rich hidden universe. The Higgs portal interaction to dark fermions or vectors could proceed via a higher-dimensional operator if the mediator is heavy. This raises the question of how potential experimental anomalies or even the discovery of a related phenomenological signature at the HL-LHC and beyond can be further scrutinised to reveal a more fine-grained picture of the underlying dynamics. Can one disentangle the different possibilities for the Higgs portal? This is the central question that our work addresses.

Given the currently observed Higgs properties, this is a difficult task when direct decays of the 125 GeV Higgs boson are open, as they are already well constrained by measurements sensitive to invisible Higgs decays. On the other hand, for heavier states above the invisible decay threshold of half the Higgs mass, it is by no means clear that there is indeed a positive answer, as there are very few collider handles to scrutinise singlet extensions. 

We studied this question through the lens of an integrated Higgs phenomenology programme bridging across present and future collider experiments, from the HL-LHC over the FCC-ee to FCC-hh. Only through a combination of different relevant observables across this ambitious, long-term programme can we maximise the information necessary to comprehensively probe the Higgs sector. We find that if a discovery (or evidence) of a Higgs portal above threshold can be obtained, there are excellent prospects for discerning the nature of the portal interactions by pitting different measurement strands at present and future colliders against each other. 

In particular, precision measurements of the Higgs couplings at FCC-ee and direct measurements of the Higgs self-coupling at FCC-hh are effective at separating out the marginal portal from the dimension-5 and dimension-6 operators to hidden sector fermions and vectors, respectively, that we study here as representative examples. Shifts in the $W$ mass could also be determined extremely precisely and provide complementary sensitivity despite the Higgs portal contributing at two-loop (which we approximate by one-loop matching with one-loop running). Finally, the direct search for missing energy in weak boson fusion final states consisting of two jets plus missing transverse energy could distinguish between Higgs portals in the different relative rates predicted by their energy growth when going from HL-LHC to FCC-hh. The FCC-hh could furthermore separate out their kinematics in differential distributions of this process.    

Although some degeneracy remains in certain regions of parameter space, we conclude that sensitivity towards a hidden-sector spectrum is promising even for such a conservative scenario as difficult to probe as a Higgs portal to a $\mathbb{Z}_2$-symmetric hidden sector above threshold. The comprehensive integrated programme of electroweak and Higgs physics across FCC-ee and FCC-hh, together with HL-LHC, will be necessary should evidence for a Higgs portal emerge. It is yet another example of how precision and high-energy explorations provide complementary information to obtain a deeper understanding, in this case of the connection between the electroweak scale and a possible hidden-sector universe.

%%%%%%%%%%%%%%%%%%%%%%%%%%%%%%%%%%%%%%%%%%%%%%%%%%%%%
\subsection*{Acknowledgments}
%%%%%%%%%%%%%%%%%%%%%%%%%%%%%%%%%%%%%%%%%%%%%%%%%%%%%
TY is supported by the United Kingdom's Science and Technology Facilities Council (STFC) grant ST/X000753/1. MD and VM are supported by KCL NMES faculty studentships. We acknowledge the use of the King's College London e-Research CREATE High Performance Computing facilities, which provided computational resources used in this work.

%%%%%%%%%%%%%%%%%%%%%%%%%%%%%%%%%%%%%%%%%%%%%%%%%%%%%
\appendix
%%%%%%%%%%%%%%%%%%%%%%%%%%%%%%%%%%%%%%%%%%%%%%%%%%%%%
\section{Partial decay widths of portal interactions}
\label{sec:decay_widths}
%%%%%%%%%%%%%%%%%%%%%%%%%%%%%%%%%%%%%%%%%%%%%%%%%%%%%
Sufficiently light singlets coupled to the Higgs through the portal interaction contribute to the Higgs decay width, with partial widths given at tree level by
\newcommand{\csqrt}[1]{{#1}}
\begin{align}
    \Gamma_{h}^{S} &=  \frac{\lambda_{S}^2 v^2}{8\pi m_h} \beta_S, \label{eq:GammaH-RSS} \\[6pt]
    \Gamma_{h}^{\psi} &= \frac{\cpsi^2 v^2 m_h}{8\pi \Lambda^2}  \beta_\psi^3, \\[6pt]
    \Gamma_{h}^{X} &= \frac{\cff^2 v^2}{32 \pi \Lambda^4 m_h^3} (3 + 2 \beta_X^2 + 3 \beta_X^4) \beta_X,
\end{align}
with $\beta_i = \sqrt{1-4 m_h^2 / m_i^2}$ with the corresponding singlet mass $m_i$.

%%%%%%%%%%%%%%%%%%%%%%%%%%%%%%%%%%%%%%%%%%%%%%%%%%%%%
\section{Renormalisation and radiative corrections}
\label{sec:renormalisation}
\begin{figure}[!t]
    \begin{center}
    \begin{tikzpicture}[scale=0.7, transform shape]
      \begin{feynman}
      
        \vertex (i);
        \node[shape=rectangle, fill=red, minimum size=1mm, inner sep=0pt] (v1) at (1.0, 0) {};
        \node[shape=rectangle, fill=red, minimum size=1mm, inner sep=0pt] (v2) at (3.0, 0) {};
        \vertex [right=1.0 of v2] (f);

        \diagram* {
          (i) -- [scalar] (v1),
          (v2) -- [scalar] (f),
        };

        \draw [solid] (v1) arc [start angle=180, end angle=0, radius=1];
        \draw [solid] (v2) arc [start angle=0, end angle=-180, radius=1];
      \end{feynman}

        \node[shape=rectangle, fill=red, minimum size=1mm] at (v1) {};
        \node[shape=rectangle, fill=red, minimum size=1mm] at (v2) {};
        
    \end{tikzpicture}
    \hspace{1cm}
    \begin{tikzpicture}[scale=1.0, transform shape]
      \begin{feynman}

        \vertex (i) at (0, 0);
        \vertex (f) at (3, 0);
        
        \node[shape=rectangle, fill=red, minimum size=1mm, inner sep=0pt] (v) at (1.5, 0) {};

        \diagram* {
          (i) -- [scalar] (v),
          (v) -- [scalar] (f),
        };

        \draw [solid] (v) arc [start angle=-90, end angle=270, radius=0.75];
        
      \end{feynman}

       \node[shape=rectangle, fill=red, minimum size=2mm, inner sep=0pt] at (v) {};
       
    \end{tikzpicture}
    \end{center}
    %%%%%%%%%%%%%%%%%%%%%%%%%%%%%%%%%%%%%%%%%%%%%%%%%%%%%%%%%%%%%%%%%%%%%%%%%%%%
    \centering
    \small
    \text{(a) Higgs two-point function}
    \vspace{6pt}
    %%%%%%%%%%%%%%%%%%%%%%%%%%%%%%%%%%%%%%%%%%%%%%%%%%%%%%%%%%%%%%%%%%%%%%%%%%%%
    \begin{center}
    \begin{tikzpicture}[scale=0.7, transform shape]
    \begin{feynman}
    
      \vertex (h1) at (2.0,  1.2);
      \vertex (h2) at (2.0, -1.2);
      \vertex (h3) at (-2.0,  0.0);
    
      \vertex (v4) at (0.5, 0.0);
      \vertex (v3) at (-1.0, 0.0);
    
      \diagram*{
        (h1) -- [scalar] (v4),
        (h2) -- [scalar] (v4),
        (v3) -- [scalar] (h3),
        (v4) -- [solid, half left, looseness=1.7] (v3),
        (v3) -- [solid, half left, looseness=1.7] (v4),
      };
    
    \end{feynman}
    
      \node[shape=rectangle, fill=red, minimum size=1mm] at (v3) {};
      \node[shape=rectangle, fill=red, minimum size=1mm] at (v4) {};
    
    \end{tikzpicture}
    \hspace{2em}
    \begin{tikzpicture}[scale=0.7, transform shape]
    \begin{feynman}
    
      % External legs
      \vertex (h1) at (-2.0,  1);
      \vertex (h2) at (2.0, 1);
      \vertex (h3) at (2.0,  -1.0);
    
      % Vertices
      \vertex (v4) at (0, 0.5);
      \vertex (v3) at (0.5, -1.0);
    
      \diagram*{
        (h1) -- [scalar] (v4),
        (h2) -- [scalar] (v4),
        (v3) -- [scalar] (h3),
        (v4) -- [solid, half left, looseness=1.7] (v3),
        (v3) -- [solid, half left, looseness=1.7] (v4),
      };
    
    \end{feynman}
    
      \node[shape=rectangle, fill=red, minimum size=1mm] at (v4) {};
      \node[shape=rectangle, fill=red, minimum size=1mm] at (v3) {};
    
    \end{tikzpicture}
    \hspace{2em}
    \begin{tikzpicture}[scale=0.7, transform shape]
    \begin{feynman}
    
      \vertex (h1) at (-2.0,  0);
      \vertex (h2) at (2.0, -1.5);
      \vertex (h3) at (2.0,  1.5);
    
      \vertex (v1) at (-1, 0.0);
      \vertex (v2) at (0.5, -0.86);
      \vertex (v3) at (0.5, 0.86);
      \vertex (v4) at (1, 0.0);
    
      \diagram*{
        (h1) -- [scalar] (v1),
        (h2) -- [scalar] (v2),
        (h3) -- [scalar] (v3),
        (v1) -- [solid, half right, looseness=1.7] (v4),
        (v4) -- [solid, half right, looseness=1.7] (v1),
      };
    
    \end{feynman}
    
      \node[shape=rectangle, fill=red, minimum size=1mm] at (v1) {};
      \node[shape=rectangle, fill=red, minimum size=1mm] at (v2) {};
      \node[shape=rectangle, fill=red, minimum size=1mm] at (v3) {};
    
    \end{tikzpicture}
    \end{center}
    \centering
    \small
    \text{(b) Higgs three-point function}
    \caption{1PI topologies of the Higgs portal contributions to the Higgs (a) two-point and (b) three-point functions at one-loop.}
    \label{fig:feynman-diagrams}
\end{figure}
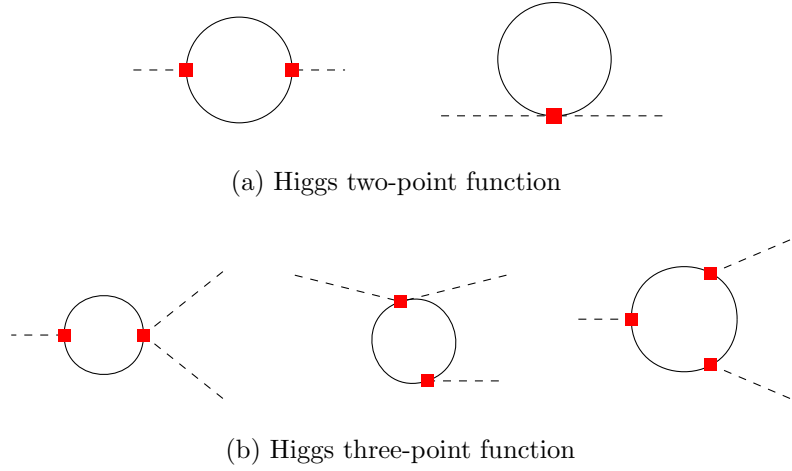

%%%%%%%%%%%%%%%%%%%%%%%%%%%%%%%%%%%%%%%%%%%%%%%%%%%%%
We compute the radiative corrections presented as indirect effects in \cref{sec:pheno} in a loop expansion, and in particular for the higher-dimensional Higgs portal operators not in an operator-dimension expansion. This ensures a consistent expansion and is appropriate since the observables can only be probed in a regime where $v/\Lambda$ is not a good expansion parameter for the operator dimension expansion. We therefore do not need the complete basis of EFT operators generated at a certain mass dimension, in particular for the effective portals, but merely all operators contributing at one-loop. The topologies of the relevant one-loop contributions to Higgs correlation functions are shown in \cref{fig:feynman-diagrams}.
We work in a mixed renormalisation scheme, following Ref.~\cite{Denner:1991kt} for the SM parameters in the on-shell (OS) scheme and parameter-renormalised tadpole scheme (PRTS), while other Wilson coefficients are renormalised in \MSbar (following the conventional approach of HEFT, see e.g.\@, Ref.~\cite{Herrero:2021iqt}). As the SM singlets only couple to the Higgs, the only one-loop contributions are to the Higgs $n$-point functions. The EFT counterterm amplitudes for the two- and three-point functions read

\begin{align}
    \delta\Sigma^\mathrm{EFT}_{hh}(p^2) &= 
        + \delta c_H\, v^2\, p^2
        + \delta a_{\Box\Box}\, v^2\, p^4 \; , \\
    \delta\Gamma^\mathrm{EFT}_{hhh}(p_1,p_2,p_3) &=
        \left(\delta c_H + \delta a_2\right) v \sum_{i=1}^{3} p_i^2
        + \left(\delta a_{\Box\Box} + \delta a_4\right) v \sum_{i=1}^{3} p_i^4
        - 6\,\delta c_6\, v^3 \; .
\end{align}

We extract the EFT counterterms by expanding the Higgs two- and three-point functions $\Sigma_{hh}$ and $\Gamma_{hhh}$, respectively, in external momenta,
\begin{align}
    \Sigma_{hh}^\mathrm{div}(p^2) &= d_{hh}^{(0)} + d_{hh}^{(2)}\, p^2 + d_{hh}^{(4)}\, p^4 \, ,\label{eq:sigmadiv} \\[4pt]
    \Gamma_{hhh}^\mathrm{div}(p_1,p_2,p_3) &= d_{hhh}^{(0)} + d_{hhh}^{(2)} \sum_{i=1}^{3} p_i^2 + d_{hhh}^{(4)} \sum_{i=1}^{3} p_i^4 \; . \label{eq:Ghhhdiv}
\end{align}
The EFT counterterms can then be mapped onto this as
\begin{equation}
\begin{gathered}
    \delta c_H = -\frac{d_{hh}^{(2)}}{v^2} \, , \quad
    \delta a_{\Box\Box} = -\frac{d_{hh}^{(4)}}{v^2} \, , \quad
    \delta c_6 = -\frac{d_{hhh}^{(0)}}{6\, v^3} \, , \\[4pt]
    \delta a_2 = \frac{d_{hhh}^{(2)}}{v} - \delta c_H \, , \quad
    \delta a_4 = \frac{d_{hhh}^{(4)}}{v} - \delta a_{\Box\Box} \, .
\end{gathered}
\end{equation}
With this, in the OS+PRTS the counterterms are then determined as
\begin{align}
    \delta M_h^2 &= \Re\Big[\Sigma_{hh}(M_h^2) + \delta c_H\, v^2 M_h^2 + \delta a_{\Box\Box}\, v^2 M_h^4\Big] \, , \\[4pt]
    \delta Z_h &= -\Re\left[\Sigma_{hh}'(M_h^2) + \delta c_H\, v^2 + 2\,\delta a_{\Box\Box}\, v^2 M_h^2\right] \, , \label{eq:dZH} \\[4pt]
    \delta t_h &= - T \, ,
\end{align}
where $T$ is the Higgs one-point function. There are no contributions to other SM counterterms at one loop. The new states are singlets under $\mathrm{SU}(2)$ and couple only through $H^\dagger H$, so they do not enter any one-particle-irreducible gauge-boson or fermion self-energy at this order, and the tadpole contributions that would otherwise feed into these through the $WWh$ and $t\bar{t}h$ vertices are removed by the PRTS condition $\delta t_h = -T$. In particular, the vacuum expectation value is not renormalised independently in the OS scheme employed here but is determined by $v = 2 M_W s_W / e$, so that $\delta v$ receives no new-physics contribution at this order.

The three portals differ in their dependence on the renormalisation scale $\mu$. 
The marginal portal is renormalisable and, as noted in \cref{sec:portals}, does not require further ultraviolet completion: the divergences of the Higgs $n$-point functions have the form of parameters already present in the Lagrangian, so that every counterterm can be fixed by an on-shell condition. These conditions are imposed at physical points and make no reference to $\mu$, and the scale dependence is absorbed into $\delta M_h^2$ and $\delta t_h$ together with the poles. The renormalised amplitudes are then manifestly
independent of the renormalisation scale: the divergence of $\Sigma^S_{hh}$ is momentum independent, so that $\delta Z_h^{S,\fin}$ involves only $\PVDBz$, and in $\hat{\Gamma}^S_{hhh}$ the logarithms cancel in the combination $2 \sum_i \PVBz(p_i^2) - 6\,\PVBz(m_h^2)$, where $\PVDBz$ and $\PVBz$ are defined in \cref{sec:oneloop-results}.

No such choice is available for the effective portals. The divergence of
$\Sigma_{hh}$ now carries powers of $p^2$, so that $c_H$ and, in the vector case, $a_{\Box\Box}$ are required in addition,\footnote{In SMEFT counting $a_{\Box\Box}$ is a dimension-eight structure. This is consistent with the loop expansion adopted here, in which the operators that appear are determined by the one-loop divergences rather than fixed in advance by their mass dimension.} together with $c_6$ and $a_{2,4}$ in the three-point function. Neither is fixed by an on-shell condition: $c_H$ enters $\Sigma_{hh}$ in the same combination as $\delta Z_h$ and cannot be separated from it there, while the conditions on the Higgs mass and on the residue of the propagator are already exhausted. These coefficients are therefore renormalised in \MSbar, which leaves an explicit $\log \mu^2$ in the renormalised amplitudes.\footnote{The renormalised self-energies given below are nonetheless $\mu$ independent in the scalar and fermion cases, the logarithms cancelling in $\PVBz(p^2) - \PVBz(m_h^2)$, whereas the vector-portal expression retains a contribution $\propto (m_h^2 - p^2)^2$: the on-shell conditions supply two subtractions, one fewer than the quartic momentum dependence of the vector-portal divergence requires. The scale dependence of the coupling modifications resides in $\delta Z_h$ and is not visible in $\hat{\Sigma}_{hh}$.}

Explicitly, for the Higgs wavefunction renormalisation we find
\begin{equation}
  \mu^2 \frac{\dd\, \delta Z_h^{\psi,\fin}}{\dd \mu^2}
    = -\frac{\cpsi^2 v^2}{8\pi^2\Lambda^2} \, , \qquad
  \mu^2 \frac{\dd\, \delta Z_h^{X,\fin}}{\dd \mu^2}
    = \frac{\cff^2 v^2}{2\pi^2\Lambda^4}\left(3 m_X^2 - m_h^2\right) \, ,
  \label{eq:muderiv}
\end{equation}
and analogous non-vanishing derivatives for $\hat{\Gamma}_{hhh}$. The \MSbar coefficients run as
\begin{equation}
    \mu^2 \frac{\dd\, c_H}{\dd \mu^2} = -\frac{d_{hh}^{(2)}}{v^2} \, , \qquad
    \mu^2 \frac{\dd\, a_{\Box\Box}}{\dd \mu^2} = -\frac{d_{hh}^{(4)}}{v^2} \, ,
    \label{eq:cHrun}
\end{equation}
with $d_{hh}^{(2)}$ and $d_{hh}^{(4)}$ the coefficients of $p^2$ and $p^4$ in $\Sigma_{hh}^\mathrm{div}$ defined in \cref{eq:sigmadiv}, so that the explicit scale dependence of \cref{eq:muderiv} is cancelled by that of the Wilson coefficients,
\begin{equation}
    \mu^2 \frac{\dd\, \Delta\kappa_h}{\dd \mu^2}
    = \frac{1}{2} \left( d_{hh}^{(2)} + 2 m_h^2\, d_{hh}^{(4)} \right)
    + \frac{1}{2}\, \mu^2 \frac{\dd\, \delta Z_h^{\fin}}{\dd \mu^2} = 0 \, .
    \label{eq:mucancel}
\end{equation}
For the fermion portal $d_{hh}^{(4)} = 0$ and \cref{eq:cHrun} reduces to $\mu^2 \dd c_H / \dd \mu^2 = - \cpsi^2 / (8\pi^2\Lambda^2)$. The coupling modification
\begin{equation}
  \Delta\kappa_h(\mu) = -\frac{1}{2}
    \left[ c_H(\mu) + 2\, a_{\Box\Box}(\mu)\, m_h^2 \right] v^2
    + \frac{1}{2}\, \delta Z_h^{\fin}(\mu)
  \label{eq:kappah-full}
\end{equation}
is independent of $\mu$ once these coefficients are treated as parameters of the effective theory. The residual scale dependence is therefore not a theoretical uncertainty in the usual sense. A tree-level insertion of $\partial_\mu (H^\dagger H) \partial^\mu (H^\dagger H)$ modifies $\kappaZZH$ in the same way and at the same order in $1/\Lambda$ as the loop, so that only the combination in \cref{eq:kappah-full} is observable. One could fix $c_H$ from measurement, but since the coupling shift is universal this would turn $\kappaZZH$ into an input and forfeit it as a prediction. We therefore fix the \MSbar coefficients by assumption.

Throughout we set $c_H(\mu_0) = a_{\Box\Box}(\mu_0) = c_6(\mu_0) = a_{2}(\mu_0) = a_{4}(\mu_0) = 0$ at $\mu_0 = 2 m_h$, the di-Higgs threshold at which the three-point functions are evaluated, so that at the scale of the observables the coupling modifications arise from the portal interaction alone. This places the three portals on an equal footing since the marginal portal is UV complete by itself and hence the Wilson coefficients vanish identically. Imposing the same condition on the effective portals separates the effect of the portal operator from deformations of the Higgs sector that the UV completion may generate in addition. This condition holds at the scale $\mu_0$ only, and the coefficients are non-vanishing at different scales. We stress that this is an assumption about the UV completion and not a convention, since a completion generating $c_H(\mu_0) \neq 0$ would shift $\kappaZZH$ at the same order as the loop contribution from the portal itself. Relaxing this assumption introduces further free, non-zero Wilson coefficients that the two coupling modifications considered here cannot constrain. Discriminating the portals would then require observables sensitive to the momentum dependence that distinguishes the corresponding operators, such as the $m_{hh}$ distribution in di-Higgs production, which we leave to future work.

% ===========================================================================
\subsection{One-loop amplitudes}
\label{sec:oneloop-results}
% ===========================================================================
We collect here the renormalised Higgs two- and three-point functions for the scalar, fermion and vector portals, expressed in terms of the scalar one-loop integrals
\begin{align}
    \PVAz(m_0^2) &= \frac{\mu^{4-d}}{i\pi^{d/2}} \int \d^d q \,
        \frac{1}{q^2 - m_0^2} \, , \\
    \PVBz(p^2, m_0^2, m_1^2) &= \frac{\mu^{4-d}}{i\pi^{d/2}} \int \d^d q \,
        \frac{1}{(q^2 - m_0^2)\big((q+p)^2 - m_1^2\big)} \, , \\
    \PVCz(p_1^2, p_2^2, p_{12}^2, m_0^2, m_1^2, m_2^2) &= \frac{1}{i\pi^2}
        \int \d^4 q \, \frac{1}{(q^2 - m_0^2)\big((q+p_1)^2 - m_1^2\big)
        \big((q+p_1+p_2)^2 - m_2^2\big)} \, ,
\end{align}
understood here as the finite pieces in the limit $d\to4$, and $\PVDBz(p^2,m_0^2,m_1^2) \equiv \partial\PVBz (p^2,m_0^2,m_1^2)/\partial p^2$. The OS-renormalised Higgs self-energies read
\begin{align}
    \hat{\Sigma}_{hh}^{S}  (p^2) &= \frac{2\lambda_S^2 v^2}{16\pi^2}
    \Big[ \PVBz(p^2,m_S^2,m_S^2) -  \Re \PVBz(m_h^2,m_S^2,m_S^2) \nn \\
    &\qquad\qquad + (m_h^2-p^2)\,\Re\PVDBz(m_h^2,m_S^2,m_S^2) \Big] \, , \\[4pt]
    \hat{\Sigma}_{hh}^{\psi} (p^2) &= \frac{\cpsi^2 v^2}{8\pi^2 \Lambda^2}
    \Big[ (p^2-4m_\psi^2)\big(\PVBz(p^2,m_\psi^2,m_\psi^2)
        - \Re\PVBz(m_h^2,m_\psi^2,m_\psi^2)\big) \nn \\
    &\qquad\qquad - (m_h^2-4m_\psi^2)(p^2-m_h^2)\,
        \Re\PVDBz(m_h^2,m_\psi^2,m_\psi^2) \Big] \, , \\
    \hat{\Sigma}_{hh}^{X} (p^2)  &= \frac{4\cff^2 v^2}{16\pi^2 \Lambda^4 }\Big[
      \left(p^4 - 4 m_X^2 p^2 + 6 m_X^4\right)
      \big(\PVBz(p^2, m_X^2, m_X^2)
         - \Re \PVBz (m_h^2, m_X^2, m_X^2)\big)
  \nonumber\\[6pt]
  &\hspace{4.2em}
    + \left(m_h^2 - p^2\right)^2 \Re \PVBz (m_h^2, m_X^2, m_X^2) 
  \nonumber\\[6pt]
  &\hspace{4.2em}
    + \left(m_h^2 - p^2\right)
      \left(m_h^4 - 4 m_h^2 m_X^2 + 6 m_X^4\right)
      \Re \PVDBz (m_h^2, m_X^2, m_X^2)
  \Big],
\end{align}
from which the finite Higgs wavefunction renormalisation constants entering
$\kappaZZH$ follow as
\begin{align}
    \delta Z_h^{S,\fin} &=
        -\frac{\lambda_S^2 v^2}{8\pi^2}\,
        \Re\PVDBz(m_h^2,m_S^2,m_S^2) \, , \\[4pt]
    \delta Z_h^{\psi,\fin} &=
        -\frac{\cpsi^2 v^2}{8\pi^2\Lambda^2}\Big[
        \Re\PVBz(m_h^2,m_\psi^2,m_\psi^2)
        + (m_h^2-4m_\psi^2)\,\Re\PVDBz(m_h^2,m_\psi^2,m_\psi^2)\Big] \, , \\[4pt]
    \delta Z_h^{X,\fin} &=
        -\frac{\cff^2 v^2}{4\pi^2\Lambda^4}\Big[
        - 2 \Re \PVAz (m_X^2) + 2 (m_h^2 -2 m_X^2) \Re \PVBz (m_h^2,m_X^2,m_X^2)\nn  \\[4pt]
        & \qquad \qquad\quad  + (m_h^4 - 4m_h^2 m_X^2 + 6 m_X^4)\,\Re\PVDBz(m_h^2,m_X^2,m_X^2)\Big] \, .
\end{align}
The renormalised three-point functions entering $\kappaHHH$ are 
\begin{align}
    \hat{\Gamma}_{hhh}^{S} &= \frac{\lambda_S^2 v}{16\pi^2}\Big[
        2\sum_{i=1}^{3}\PVBz(p_i^2,m_S^2,m_S^2)
        - 6\,\Re\PVBz(m_h^2,m_S^2,m_S^2)
    + 9\,m_h^2\,\Re\PVDBz(m_h^2,m_S^2,m_S^2) \Big] \nn \\
    &\quad + \frac{8 \lambda_S^3 v^3}{16\pi^2} \PVCz(p_2^2,p_3^2,p_1^2,m_S^2,m_S^2,m_S^2)  \, , \label{eq:gammaHHHS}\\[4pt]
    \hat{\Gamma}_{hhh}^{\psi} &= \frac{\cpsi^2 v}{16\pi^2\Lambda^2}\Big[
        2\sum_{i=1}^{3}\big(p_i^2-4m_\psi^2\big)\,\PVBz(p_i^2,m_\psi^2,m_\psi^2)
        + 3\big(m_h^2+8m_\psi^2\big)\,\Re\PVBz(m_h^2,m_\psi^2,m_\psi^2) \nn \\
        &\qquad\qquad + 9\,m_h^2\big(m_h^2-4m_\psi^2\big)\,\Re\PVDBz(m_h^2,m_\psi^2,m_\psi^2) \Big] \nn \\
    &\quad +\frac{4\cpsi^3 v^3 m_\psi}{16\pi^2 \Lambda^3} \Big[
        - 2\sum_{i=1}^{3}\PVBz(p_i^2,m_\psi^2,m_\psi^2)
        + \Big(\sum_{i=1}^{3}p_i^2 - 8 m_\psi^2\Big)
        \PVCz(p_2^2,p_3^2,p_1^2,m_\psi^2,m_\psi^2,m_\psi^2)
    \Big] \, ,\label{eq:gammaHHHD} \\[4pt]
    \hat{\Gamma}_{hhh}^{X} &= \frac{\cff}{16\pi^2\Lambda^2 v}\Big[\, 12\,m_X^4 \,\Big] \nn \\
&\quad + \frac{\cff^2 v}{16\pi^2\Lambda^4}\bigg[
    - 4\sum_{i=1}^{3}\Big(p_i^4 - 6m_X^2 p_i^2 + 12\,m_X^4\Big)
    - 4\Big(3m_h^2 + 2\sum_{i=1}^{3}p_i^2\Big)\PVAz(m_X^2) \nn \\
&\qquad\qquad\quad + 4\sum_{i=1}^{3}\mathcal{B}_X(p_i^2)
    - 12\,\Re\mathcal{B}_X(m_h^2)
    + 18\,m_h^2\,\Re\mathcal{B}_X'(m_h^2) \bigg] \nn \\
&\quad + \frac{\cff^3 v^3}{16\pi^2\Lambda^6}\bigg[
    8\Big(\big(\textstyle\sum_{i=1}^{3}p_i^2\big)^2 - 18\,m_X^2\sum_{i=1}^{3}p_i^2 + 96\,m_X^4\Big)
    + 32\Big(\sum_{i=1}^{3}p_i^2 - 18\,m_X^2\Big)\PVAz(m_X^2) \nn \\
&\qquad\qquad\quad - 16\sum_{i=1}^{3}\Big(\mathcal{F}(p_i^2) - m_X^2\big(\textstyle\sum_{j=1}^{3}p_j^2 - 6m_X^2\big)\Big)\PVBz(p_i^2,m_X^2,m_X^2) \nn \\
&\qquad\qquad\quad - 16\,m_X^2\sum_{i=1}^{3}\big(p_i^2-2m_X^2\big)^2\,\PVCz(p_2^2,p_3^2,p_1^2,m_X^2,m_X^2,m_X^2)
\bigg] \, ,
\label{eq:gammaHHHX}
\end{align}
where
\begin{align}
    \mathcal{F}(s) \equiv \lambda(s,m_X^2,m_X^2) + 6m_X^4 \,, \qquad
    \mathcal{B}_X(s) \equiv \mathcal{F}(s)\,\PVBz(s,m_X^2,m_X^2) \,, \qquad
    \mathcal{B}_X'(s) \equiv \frac{\dd \mathcal{B}_X}{\dd s} \,,
\end{align}
and $\lambda$ is the K\"all\'en function,
\begin{equation}
    \lambda(x,y,z) = x^2 + y^2 + z^2 - 2xy - 2yz - 2xz \, .
\end{equation}
The three-point functions are evaluated at the kinematic point $p_1^2=p_2^2=m_h^2$, $p_3^2 = p_{12}^2 = 4m_h^2$ as discussed in \cref{sec:kappalambda}.

Two structural features of these expressions underlie the correlations discussed in \cref{sec:discrimination}. First, the self-energies are quadratic in the portal coupling $c$, whereas the three-point functions contain a term cubic in it through the triangle topology, so that $\Delta\kappaZZH \propto c^2$ while $\Delta\kappaHHH$ retains a linear sensitivity to the sign of $c$. Second, the cubic term in the fermionic case is proportional to $m_\psi$, reflecting the chirality flip required by the $\bar\psi\psi$ structure, and no such suppression occurs for the marginal scalar portal.
Furthermore, the vector portal contributes at linear order in $\cff$ to the Higgs three-point function, entering through $\delta t_h$ and $\delta M_h^2$. Contracting $F_{\mu\nu}F^{\mu\nu}$ across the loop yields a factor $(1-d)$, whose $\mathcal{O}(\epsilon)$ part multiplies the pole of $\PVAz(m_X^2)$ and leaves a rational remainder. In the renormalised vertex function the $\PVAz$ terms and the poles cancel, giving the $12 m_X^4$ term above. The marginal and fermionic portals have no such $d$-dependent prefactor and receive no linear contribution.

\subsection{One-loop convergence}
\label{sec:oneloopconvergence}

The expressions in \cref{eq:gammaHHHS,eq:gammaHHHD,eq:gammaHHHX} are valid for the one-loop approximation, where contributions of different mass dimensions in the BSM scales of the respective portal interactions are encoded. Separating these contributions according to their EFT scaling enables us to stress test the parameter regions identified earlier with regard to the relevance of these higher-dimensional effects within the one-loop approximation. 

The scalar portal, as expected for a renormalisable theory, decouples for large enough exotic scalar mass. Even when its approximation as an EFT is justified, the resummation of the new physics scale $\sim m_S$ expressed in the loop order is theoretically preferred~\cite{Coleman:1973jx}. This is fundamentally different for the fermion and vector portals, as the loop and EFT expansion are mixed by construction. Relying on HEFT for the one-loop evaluation, we can identify higher-dimensional structures $\Lambda^{n<-2}$ with different kinematic dependence that arise when the interactions considered in this work are indeed dominant (i.e. competing operators are assumed to be absent, as detailed above). The relative importance of the higher-dimensional $\Lambda^{n<-2}$ terms sheds light on the convergence of the theory within our one-loop approximation, i.e. the compatibility of the loop expansion with the EFT expansion captured by it. In particular, the linear term in \cref{eq:gammaHHHX} causes the vector-portal contribution to grow as $\propto m_X^4$ as a consequence of rational terms, while the fermion-portal contribution grows logarithmically. Consequently, the nominally discoverable regions of parameter space extend towards increasingly large masses and small couplings. It is known that when a concrete renormalisable UV theory can be identified, such behaviour is avoided, with critical importance of the renormalisation scheme~\cite{Dittmaier:2021fls}. Questions of perturbative convergence, however, also arise in the renormalisable scalar portal when the coupling constraints are not strong enough to render two-loop contributions suppressed~\cite{Englert:2019eyl}.

%%%%%%%%%%%%%%%%%%%%%%%%%%%%%%%%%%%%%%%%%%%%%%%%%%%%%%%%%%%%%%%%%%%%%%%%%%%%%%%%
\begin{figure}[!t]
    \centering
    \includegraphics[width=\linewidth]{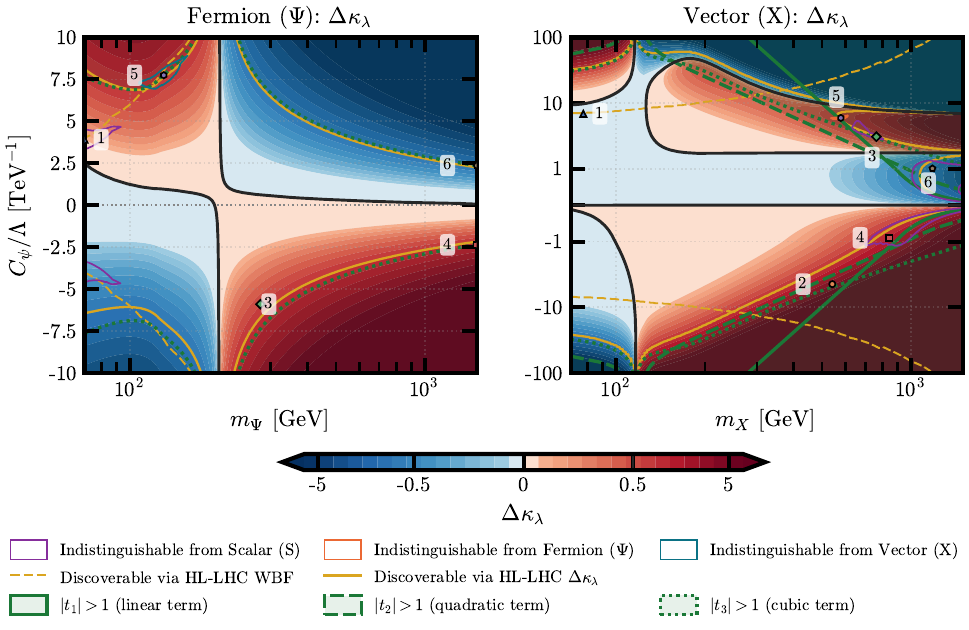}
    \caption{Contour plot in the plane of mass vs coupling showing $\Delta \kappaHHH$ for the fermionic (left) and vector (right) portals. The points shown correspond to the benchmark points in \cref{tab:benchmarks}. The colour coding corresponds to a signed logarithmic scale centred around 0, with positive (negative) values in increasingly darker shades of red (blue). The contours along which the sign of the coupling modifiers changes are denoted by a solid black line. The discoverable and (in)distinguishable regions from \cref{fig:confusion_discovery_hllhc} are overlaid as unshaded contours following the same colour coding. The green contours delimited by a solid, dashed and dotted green line correspond to the region at which the linear, quadratic and cubic terms become greater than 1 and so potentially sensitive to higher-order corrections. }
    \label{fig:kappa_convergence}
\end{figure}
%%%%%%%%%%%%%%%%%%%%%%%%%%%%%%%%%%%%%%%%%%%%%%%%%%%%%%%%%%%%%%%%%%%%%%%%%%%%%%%%

To illustrate this specifically for the EFT scenarios, we note that the expression for $\Delta \kappa_\lambda$ is a third-degree polynomial 
\begin{equation}
\Delta\kappa_\lambda
=
\frac{\Re\hat{\Gamma}_{hhh}
-\Re\hat{\Gamma}_{hhh}^{\rm SM}}
{\Gamma_{hhh}^{\rm tree}}
=
a_1 C+a_2C^2+a_3C^3,
\end{equation}
where $C$ is the relevant dimensionful Wilson coefficient. We can now define regions in which any individual contribution satisfies
\begin{equation}
|t_n| = |a_n C^n|\gtrsim1
\end{equation}
as potentially susceptible to higher loop or EFT-expansion order. The corresponding regions of parameter space are illustrated in \cref{fig:kappa_convergence}. They show $\Delta\kappaHHH$ in the mass–coupling plane for the fermion (left) and vector (right) portals. The points indicate the benchmark scenarios defined in \cref{tab:benchmarks}. The colour coding uses a signed logarithmic scale centred at zero, with positive (negative) values represented by increasingly dark shades of red (blue). The contours along which the coupling modifier changes sign are indicated by solid black lines. The discoverable and (in)distinguishable regions from \cref{fig:confusion_discovery_hllhc} are overlaid as unshaded contours following the same colour coding. The solid, dashed, and dotted green contours delimit the regions in which the linear, quadratic, and cubic contributions, respectively, satisfy $|t_n|=1$. Points beyond these contours are therefore potentially susceptible to higher-order corrections but not necessarily unphysical.

As alluded to above, the vector portal shows these effects most obviously; the contributions across the EFT expansion for an ${\cal{O}}(1)$ correction to the Higgs self-coupling (which is conservative given the current experimental constraints) are comparable to $m_X\simeq 800~\text{GeV}$. 
They are considerably milder for the fermion portal; the quadratic contribution exceeds unity only for couplings outside the plotted range. Crucially, several of the benchmark points contributing to the indistinguishable regions in \cref{fig:confusion_discovery_hllhc} lie within the region where additional theoretical input will be needed if a discovery is made.

%%%%%%%%%%%%%%%%%%%%%%%%%%%%%%%%%%%%%%%%%%%%%%%%%%%%%%%%%%%%%%%%%%%%%%%%%%%%%%%%
\section{Weak boson fusion analysis}
\label{sec:wbf-appendix}
%%%%%%%%%%%%%%%%%%%%%%%%%%%%%%%%%%%%%%%%%%%%%%%%%%%%%%%%%%%%%%%%%%%%%%%%%%%%%%%%
Here we share some further details on the analysis done to obtain the weak boson
fusion constraints of \cref{sec:wbf}.

Signal and background samples are generated with \madgraph at leading order
using our own \feynrules model, showered with \pythia (MLM matched for the QCD
$V+jj$ samples) and passed through \delphes, with the CMS detector card for LHC
studies and the FCC-hh scenario II card for FCC-hh~\cite{FCChhPhysicsPerformance}. The background
comprises QCD and electroweak $Z(\nu\nu)jj$ and $W(\ell\nu)jj$, summed after
selection.

To improve the signal-to-background ratio, we enforce the following cuts on the
kinematic distributions. We require at least two jets per event with a transverse
momentum $p_T > \SI{50}{\GeV}$ within pseudo-rapidity $|\eta| < 4.7$; events with
additional jets are retained. Events containing an isolated lepton above
\SI{10}{\GeV} are vetoed, which removes the $W(\ell\nu)jj$ background whenever
the lepton is reconstructed. The two hardest jets are taken as the tagging pair
and are required to lie in opposite hemispheres, $\eta_1\eta_2 < 0$, and no
further jet above \SI{30}{\GeV} may fall in the rapidity interval between them.
We impose a minimum angular separation between the missing momentum and the jets,
$|\Delta\phi(\met, j_i)| > 0.5$, and a maximum separation between the tagging
jets, $\Delta\phi(j_1,j_2) < 2.2$.

%%%%%%%%%%%%%%%%%%%%%%%%%%%%%%%%%%%%%%%%%%%%%%%%%%%%%
\begin{figure}
    \centering
    \includegraphics[width=\linewidth]{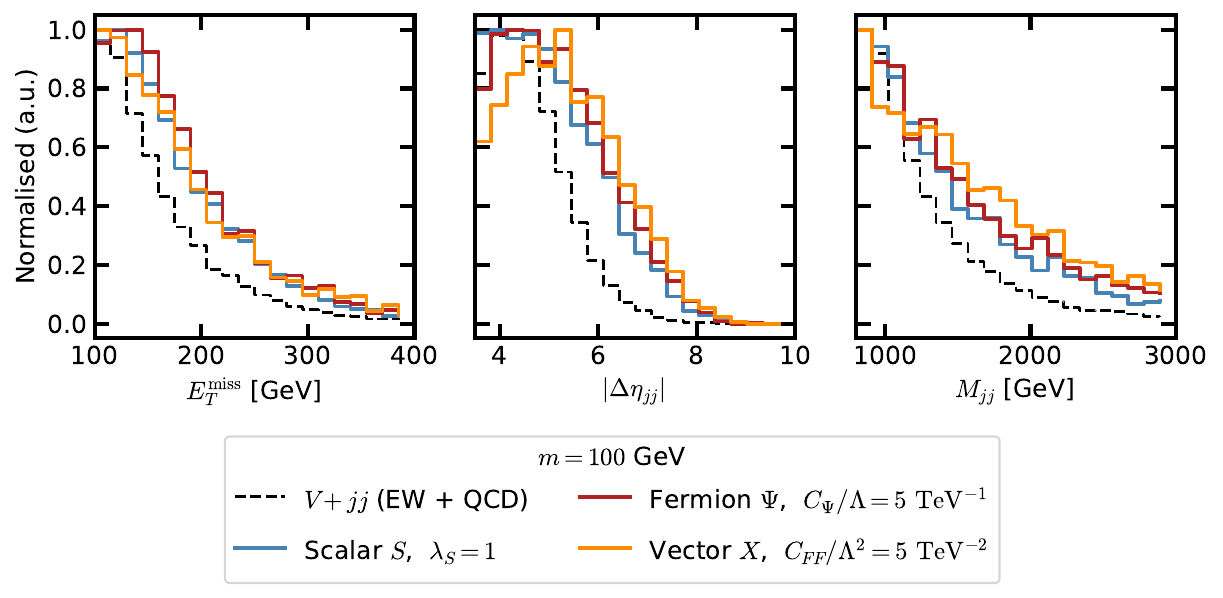}
    \caption{Normalised distributions of the most relevant kinematic variables for weak boson fusion production of two invisible dark-sector particles through the scalar (blue), fermionic (red), or vector (orange) Higgs portal, for representative coupling values and a light dark-sector mass of $M=\SI{100}{\giga\electronvolt}$ at HL-LHC. The black dashed line corresponds to the Standard Model background from vector-boson production in association with jets, with the vector bosons decaying into neutrinos.}
    \label{fig:kinematic_distributions_generic}
\end{figure}
%%%%%%%%%%%%%%%%%%%%%%%%%%%%%%%%%%%%%%%%%%%%%%%%%%%%%

We also require the jets to have a large rapidity gap, $|\detajj| > 4.2$. The
rapidity gap is the characteristic WBF
signature~\cite{Cahn:1983ip,Dokshitzer:1987nc,Dokshitzer:1991he,Rainwater:1996ud,%
Rainwater:1999sd,Plehn:1999xi}: the tagging jets follow the incoming quarks and
emerge at small angles to the two beams, whereas the QCD background populates the
whole rapidity range. Additionally, we require that $\met > \SI{180}{\GeV}$. The
mediators are produced through an off-shell Higgs and are heavy, so the invisible
system carries substantial transverse momentum. As a last step, we optimise the
$M_{jj}$ cut to be $M_{jj} \gtrsim 1$ TeV in order to maximise $S/\sqrt{B}$. These distributions show a significant separation between SM background and
signal for large $\met$, $M_{jj}$ and $|\detajj|$, motivating the cuts that
select this kinematic region. The full cutflow is summarised for one benchmark point in \cref{tab:cutflow_lhc}.

%%%%%%%%%%%%%%%%%%%%%%%%%%%%%%%%%%%%%%%%%%%%%%%%%%%%%
\begin{table}[!t]
\centering
\caption{HL-LHC cut-flow for the three portals at $m = \SI{100}{\GeV}$, with
$\lambda_S = 1$, $\cpsi/\Lambda = \SI{5}{\per\TeV}$ and $\cff/\Lambda^2 = \SI{5}{\per\TeV\squared}$, against the summed $V+jj$ background.
Entries are expected event counts at \SI{3}{\per\atto\barn}.}
\label{tab:cutflow_lhc}
\vspace{0.25cm}
\begin{tabular}{lrrrr}
\toprule
Cut & $S$ (\Sc) & $S$ (\Fm) & $S$ (\Vc) & $B$ \\
\midrule
No cuts                                  & $5.06\cdot10^{4}$ & $5.80\cdot10^{4}$ & $3.05\cdot10^{4}$ & $3.70\cdot10^{7}$ \\
$\ge 2$ jets, $p_T > \SI{50}{\GeV}$, $|\eta| < 4.7$
                                         & $2.15\cdot10^{4}$ & $2.53\cdot10^{4}$ & $1.31\cdot10^{4}$ & $2.89\cdot10^{7}$ \\
Lepton veto, $p_T > \SI{10}{\GeV}$       & $2.15\cdot10^{4}$ & $2.53\cdot10^{4}$ & $1.31\cdot10^{4}$ & $2.29\cdot10^{7}$ \\
Opposite hemispheres, $\eta_1\eta_2 < 0$ & $1.65\cdot10^{4}$ & $2.01\cdot10^{4}$ & $1.07\cdot10^{4}$ & $2.05\cdot10^{7}$ \\
Central-jet veto, $p_T > \SI{30}{\GeV}$  & $1.17\cdot10^{4}$ & $1.41\cdot10^{4}$ & $7.38\cdot10^{3}$ & $9.50\cdot10^{6}$ \\
\midrule
$\Delta\phi(j_1,j_2) < 2.2$              & $7.51\cdot10^{3}$ & $9.33\cdot10^{3}$ & $4.95\cdot10^{3}$ & $4.91\cdot10^{6}$ \\
$|\Delta\phi(\met,j)| > 0.5$             & $7.21\cdot10^{3}$ & $8.96\cdot10^{3}$ & $4.70\cdot10^{3}$ & $4.57\cdot10^{6}$ \\
$|\detajj| > 4.2$                        & $4.17\cdot10^{3}$ & $5.56\cdot10^{3}$ & $3.35\cdot10^{3}$ & $3.25\cdot10^{6}$ \\
$\met > \SI{180}{\GeV}$                  & $1.04\cdot10^{3}$ & $1.56\cdot10^{3}$ & $1.04\cdot10^{3}$ & $6.68\cdot10^{5}$ \\
$M_{jj} > \SI{1005}{\GeV}$                 & $9.06\cdot10^{2}$ & $1.38\cdot10^{3}$ & $9.40\cdot10^{2}$ & $5.41\cdot10^{5}$ \\
\midrule
$S/\sqrt{B}$                             & $1.2$ & $1.9$ & $1.3$ & \\
\bottomrule
\end{tabular}
\end{table}
%%%%%%%%%%%%%%%%%%%%%%%%%%%%%%%%%%%%%%%%%%%%%%%%%%%%%

For the FCC-hh we repeat the analysis at $\sqrt{s} = \SI{84}{\TeV}$ with \SI{30}{\per\atto\barn}, keeping the same cuts and only adjusting for the larger centre-of-mass energy: The jet acceptance is widened to $|\eta| < 6.0$, matching the FCC-hh detector design and recovering the tagging jets, which are produced further forward; the missing-energy threshold is raised slightly to \SI{200}{\GeV}; and the $M_{jj}$ scan is moved up to start at \SI{2300}{\GeV} since both signal and background are far harder at \SI{84}{\TeV}.

Lastly, because the Higgs-associated production cross sections for the singlets
factorise as $\sigma(m,c) = c^2\,\hat\sigma(m)$, with $m$ the singlet mass and
$c$ the portal coupling, and with all selection efficiencies depending on the
mass alone, one sample per (mediator, mass) point suffices: the coupling can be
varied by a simple reweighting, and the cut optimisation is performed once per
mass at a reference value for the coupling.

%%%%%%%%%%%%%%%%%%%%%%%%%%%%%%%%%%%%%%%%%%%%%%%%%%%%%
\begin{table}[!t]
\centering

\caption{FCC-hh cut-flow for the three portals at $m = \SI{100}{\GeV}$, with
$\lambda_S = 1$, $\cpsi/\Lambda = \SI{5}{\per\TeV}$ and $\cff/\Lambda^2 = \SI{5}{\per\TeV\squared}$, against the summed $V+jj$ background.
Entries are expected event counts at \SI{30}{\per\atto\barn}.}
\label{tab:cutflow_fcchh}
\vspace{0.25cm}
\begin{tabular}{lrrrr}
\toprule
Cut & $S$ (\Sc) & $S$ (\Fm) & $S$ (\Vc) & $B$ \\
\midrule
No cuts                                  & $9.48\cdot10^{6}$ & $1.52\cdot10^{7}$ & $2.31\cdot10^{7}$ & $2.82\cdot10^{9}$ \\
$\ge 2$ jets, $p_T > \SI{50}{\GeV}$, $|\eta| < 6$
                                         & $5.17\cdot10^{6}$ & $8.60\cdot10^{6}$ & $1.29\cdot10^{7}$ & $2.65\cdot10^{9}$ \\
Lepton veto, $p_T > \SI{10}{\GeV}$       & $5.16\cdot10^{6}$ & $8.59\cdot10^{6}$ & $1.29\cdot10^{7}$ & $1.77\cdot10^{9}$ \\
Opposite hemispheres, $\eta_1\eta_2 < 0$ & $3.50\cdot10^{6}$ & $6.23\cdot10^{6}$ & $9.83\cdot10^{6}$ & $1.54\cdot10^{9}$ \\
Central-jet veto, $p_T > \SI{30}{\GeV}$  & $1.95\cdot10^{6}$ & $3.28\cdot10^{6}$ & $4.80\cdot10^{6}$ & $4.93\cdot10^{8}$ \\
\midrule
$\Delta\phi(j_1,j_2) < 2.2$              & $1.29\cdot10^{6}$ & $2.20\cdot10^{6}$ & $3.30\cdot10^{6}$ & $2.05\cdot10^{8}$ \\
$|\Delta\phi(\met,j)| > 0.5$             & $1.20\cdot10^{6}$ & $2.05\cdot10^{6}$ & $3.04\cdot10^{6}$ & $1.80\cdot10^{8}$ \\
$|\detajj| > 4.2$                        & $9.25\cdot10^{5}$ & $1.71\cdot10^{6}$ & $2.76\cdot10^{6}$ & $1.54\cdot10^{8}$ \\
$\met > \SI{200}{\GeV}$                  & $2.54\cdot10^{5}$ & $5.32\cdot10^{5}$ & $9.00\cdot10^{5}$ & $5.16\cdot10^{7}$ \\
$\mjj > \SI{2300}{\GeV}$                 & $1.43\cdot10^{5}$ & $3.43\cdot10^{5}$ & $7.00\cdot10^{5}$ & $2.18\cdot10^{7}$ \\
\midrule
$S/\sqrt{B}$                             & $30.5$ & $73.5$ & $149.7$ & \\
\bottomrule
\end{tabular}
\end{table}
%%%%%%%%%%%%%%%%%%%%%%%%%%%%%%%%%%%%%%%%%%%%%%%%%%%%%

%%%%%%%%%%%%%%%%%%%%%%%%%%%%%%%%%%%%%%%%%%%%%%%%%%%%%
%  bibliography
%%%%%%%%%%%%%%%%%%%%%%%%%%%%%%%%%%%%%%%%%%%%%%%%%%%%% 
\bibliographystyle{JHEP}
\bibliography{references}
%%%%%%%%%%%%%%%%%%%%%%%%%%%%%%%%%%%%%%%%%%%%%%%%%%%%%   
\end{document}